\documentclass[usenatbib]{mn2e}
\usepackage{amssymb}
\usepackage{epsfig}

\usepackage{graphicx}
\newcommand{\be}{\begin{equation}}
\newcommand{\ee}{\end{equation}}
\newcommand{\bea}{\begin{eqnarray}}
\newcommand{\eea}{\end{eqnarray}}
 \newcommand{\rxte}{{\it RXTE}}
\newcommand{\rs}{R_{\rm S}}
\newcommand{\fedd}{F_{\rm Edd}}
\newcommand{\msun}{\rm{M}_{\odot}} 
 \newcommand{\source}{4U~1608--52} 
 
 \newcommand{\araa}{ARA\&A}  
 \newcommand{\apjs}{ApJS}  
 \newcommand{\apj}{ApJ}  
 \newcommand{\apjl}{ApJ}  
 \newcommand{\aap}{A\&A}  
 \newcommand{\mnras}{MNRAS}  
 \newcommand{\pasj}{PASJ} 
  \newcommand{\nat}{Nat}  
   \newcommand{\physrep}{Phys. Rep.}

\title[Neutron star mass and radius measurements in 4U 1608$-$52]{
The effect of accretion on the measurement of neutron star mass and radius  in 
the low-mass X-ray binary 4U 1608$-$52}
 
\author[J. Poutanen et al.]
{Juri Poutanen,$^{1,2}$\thanks{E-mail:juri.poutanen@gmail.com}  
Joonas N\"attil\"a,$^{1,2}$ 
Jari J. E. Kajava,$^{3,4,2}$ 
Outi-Marja Latvala,$^2$ \and
Duncan Galloway,$^{5,6}$
Erik Kuulkers$^3$
and Valery Suleimanov$^{7,8}$
\\
$^1$Tuorla Observatory, University of Turku, V\"ais\"al\"antie 20, FI-21500 Piikki\"o, Finland\\
$^2$Astronomy Division, Department of Physics, PO Box 3000, FI-90014 University of Oulu, Finland \\
$^3$European Space Astronomy Centre (ESA/ESAC), Science Operations
Department, 28691 Villanueva de la Ca\~{n}ada, Madrid, Spain\\
$^4$Nordic Optical Telescope, Apartado 474, 38700 Santa Cruz de La Palma, Spain\\
$^5$Monash Centre for Astrophysics (MoCA), Monash University, Clayton, VIC 3800, Australia\\
$^6$School of Physics and School of Mathematical Sciences, Monash University, Clayton, VIC 3800, Australia \\
$^7$Institut f\"ur Astronomie und Astrophysik, Kepler Centre for Astro and
Particle Physics,   Universit\"at T\"ubingen, Sand 1, D-72076 T\"ubingen, Germany \\ 
$^8$Astronomy Department, Kazan (Volga region) Federal University,  Kremlyovskaya str., 18, 420008 Kazan, Russia\\
}

\begin{document}
\date{Accepted. Received ; in original form }
\pagerange{\pageref{firstpage}--\pageref{lastpage}}
\pubyear{2014}

\maketitle
\label{firstpage}

 \begin{abstract}
Spectral measurements of thermonuclear (type-I) X-ray bursts from low mass X-ray binaries have been used to measure neutron star 
(NS) masses and radii. 
A number of systematic issues affect such measurements and have raised concerns as to the robustness of the methods. 
We present analysis of the X-ray emission from bursts observed from \source\ at various persistent fluxes. 
We find a strong dependence of the burst properties on the flux and spectral hardness of the persistent emission before burst. 
Bursts occurring during the low-accretion rate (hard) state exhibit evolution of the black body normalisation consistent with the theoretical predictions 
of NS atmosphere models. 
However, bursts occurring during the high-accretion rate (soft) state show roughly constant normalisation,  
which is inconsistent with the NS atmosphere models and therefore these bursts cannot be easily used to determine NS parameters. 
We analyse the hard-state burst to put the lower limit on the neutron star radius in  \source\ of 13 km (for masses 1.2--2.4~$\msun$). 
The best agreement with the theoretical NS mass-radius relations is achieved for source distances in the range 3.1--3.7 kpc. 
We expect that the radius limit will be 10 per cent lower if spectral models including rapid rotation are used instead.  
\end{abstract}
 
 \begin{keywords}
 stars: neutron -- stars: atmospheres -- X-rays: individual (\source) -- X-rays: bursts
 \end{keywords}


\section{Introduction}
\label{sec:intro}

Neutron stars (NS) are among the most compact observable objects in our Universe.
Their core densities  can exceed the nuclear  density by a factor 2--5.
This makes them interesting testbeds of physics under extreme conditions practically unattainable in the terrestrial laboratories.
However, this also means that the equation of state (EoS) of supra-nuclear matter has large uncertainties, 
because laboratory measurements are difficult \citep{LP07} 
and computations from first principles are practically impossible, because 
of the extreme complexity of multi-body nucleon interactions \citep[e.g.][]{Chamel13}. 

Measuring NS masses and radii using astronomical observations 
offer a way to constrain the EoS \citep{HPY07,LP07,Lattimer12ARNPS}. 
Recent observations of two-solar-mass pulsars \citep{Demorest10,Antoniadis13} appear to disfavour the softest EoS. 
Better constraining the EoS from observations, however, requires not only the mass, but also the radius to be determined. 
In principle,  constraints on the radii can be obtained from the measurements of the moment of inertia, 
but it is a difficult task that might take decades to complete even in the most relativistic 
system known \citep{Lyne04,oconnell04,LS05,Kramer08}.  
Thermal emission from NS potentially offer a better tool to measure their radii. 
Cooling NS in quiescent low-mass X-ray binaries (LMXBs) situated in globular clusters with known distances 
allow to determine their apparent radii, but not masses and radii independently. 
The major problem here is that these measurements give very different, mutually excluding radii for different objects 
\citep{Guillot13} and the results depend heavily on the assumed 
chemical composition and the value of the interstellar absorption \citep{LS14quie}. 
The available data offer basically one snapshot for each object and do not allow 
thorough tests of the models. 

X-ray bursters can provide tighter constraints on the mass-radius relation \citep[e.g.][]{Damen90,vP90,LvPT93}. 
First, the so called photospheric radius expansion (PRE) bursts 
are powerful enough to exceed the Eddington limit \citep{Grindlay80,Lewin84,Tawara84}, 
which therefore can be potentially measured \citep[see][]{Kuulkers03}. 
Second, cooling of the NS surface during the burst provides a large set of time-resolved spectra, that allow 
measurement of the apparent NS radius at different fluxes. 
Furthermore, each object typically shows many bursts which can be used 
for consistency checks. If the distance to the source is known, then these observations
in principle allow constraints on both the mass and the radius. 

A serious problem encountered with this approach is that the distances are  not known  
with sufficient accuracy, resulting in large error boxes elongated along 
the curve of constant Eddington temperature (see equation (\ref{eq:tedd}) in Appendix~\ref{app:A}), 
which is distance-independent \citep{SPRW11}. 
On the other hand, some of the reported measurements  \citep{OGP09,GO10,GWCO10} give 
no solutions for mass and radius for most of the parameter space resulting 
in mass-radius constraints much more accurate than the distance error would allow. 
This casts doubts on the whole approach \citep{SLB10}. 
Another problem is that the radii determined from different objects turned out to be very different, depending 
on the applied method  (the touchdown or  the cooling tail approach) 
and the bursts selected for the analysis  \citep[see][]{SPRW11}. 
The most alarming is a clear dependence of the measured  apparent radii 
on the accretion state of the object where the burst occurs, as was shown for the case of 4U~1724$-$307 by \citet{SPRW11}. 
A more extended recent study of \citet{Kajava14} 
demonstrated that X-ray burst cooling properties in 11 LMXBs
are dependent on the accretion rate and the spectral state.  

In the present paper, we concentrate on \source, which shows PRE bursts over a wide range of persistent fluxes, and in different spectral states.
This  allows us to study the cooling of the bursts at different mass accretion rates. 
The aim of this study is to understand, using \source\ as an example, which kind of bursts evolve according to the available theoretical models, 
and  how the difference in the cooling tail  behaviour impacts the NS mass and radius measurements.

\section{PRE X-ray bursts from \source}

\subsection{The companion and the distance to \source}
\label{sect:distance}

To the best of our knowledge, there are no spectroscopic measurements of the optical counterpart QX Nor to LMXB \source. 
However,  \citet{Wachter02} have detected periodic variability at the time-scale of 0.537 days, 
which they have attributed to the super-hump period, which is very close to the orbital period of the system \citep{KFM98}. 
Observations of QX Nor  in quiescence indicate an F- to G- main-sequence secondary, while 
theoretical arguments are in favour of an evolved donor \citep{Wachter02}. 
In either case, the companion is likely to be a hydrogen-rich star.

\source\ does not reside in any known globular cluster making distance to the source hard to measure. 
Nevertheless there have been several estimates by different authors using various methods.
Using X-ray bursts, \cite{Ebi87} obtained a distance $D=3.8\pm0.4$ kpc by fitting a 
theoretical model  to the observed dependence of the colour temperature on  luminosity.
\cite{N89} gave a distance of $3.6$ kpc based on comparison between the Eddington limit for helium-rich envelope and the most luminous PRE-burst observed at the time.
More recent measurements have been made by \cite{GO10} who obtained $D=5.8_{-1.9}^{+2.0}$ kpc with a lower cutoff at 3.9 kpc, 
based on the study of the interstellar extinction toward the source.
To cover all possibilities, we assume a Gaussian distribution of distances with a mean of 5.8 kpc and standard deviation of 
2~kpc on both sides.
We will also highlight what is the effect of having the cutoff at 3.9~kpc on the $M$-$R$ constraints in Sections~\ref{sec:hard} and~\ref{sec:soft}.

\subsection{Data}
 
The data from the Proportional Counter Array (PCA) \citep{JMR06} spectrometer on board  
the {\it Rossi X-ray Timing Explorer (RXTE)} 
were analysed with the {\sc heasoft} package (version 6.11.1) and 
response matrices were generated using {\sc pcarsp} (11.7.1).  
The backgrounds of PCA detectors were estimated with \verb|CM_bright_VLE| model and all the spectral data were 
fitted using {\sc xspec} 12.8.1g package \citep{Arn96}, 
assuming a  recommended 0.5 per cent systematic error \citep{JMR06}.  
In order to take low count rate bins into account, we also adopted Churazov weighting \citep{Chur96}.
All error limits were obtained using \verb|error| -task in {\sc xspec} with $1\sigma$ confidence levels.

\begin{figure*}
\centering 
\includegraphics[width=17cm]{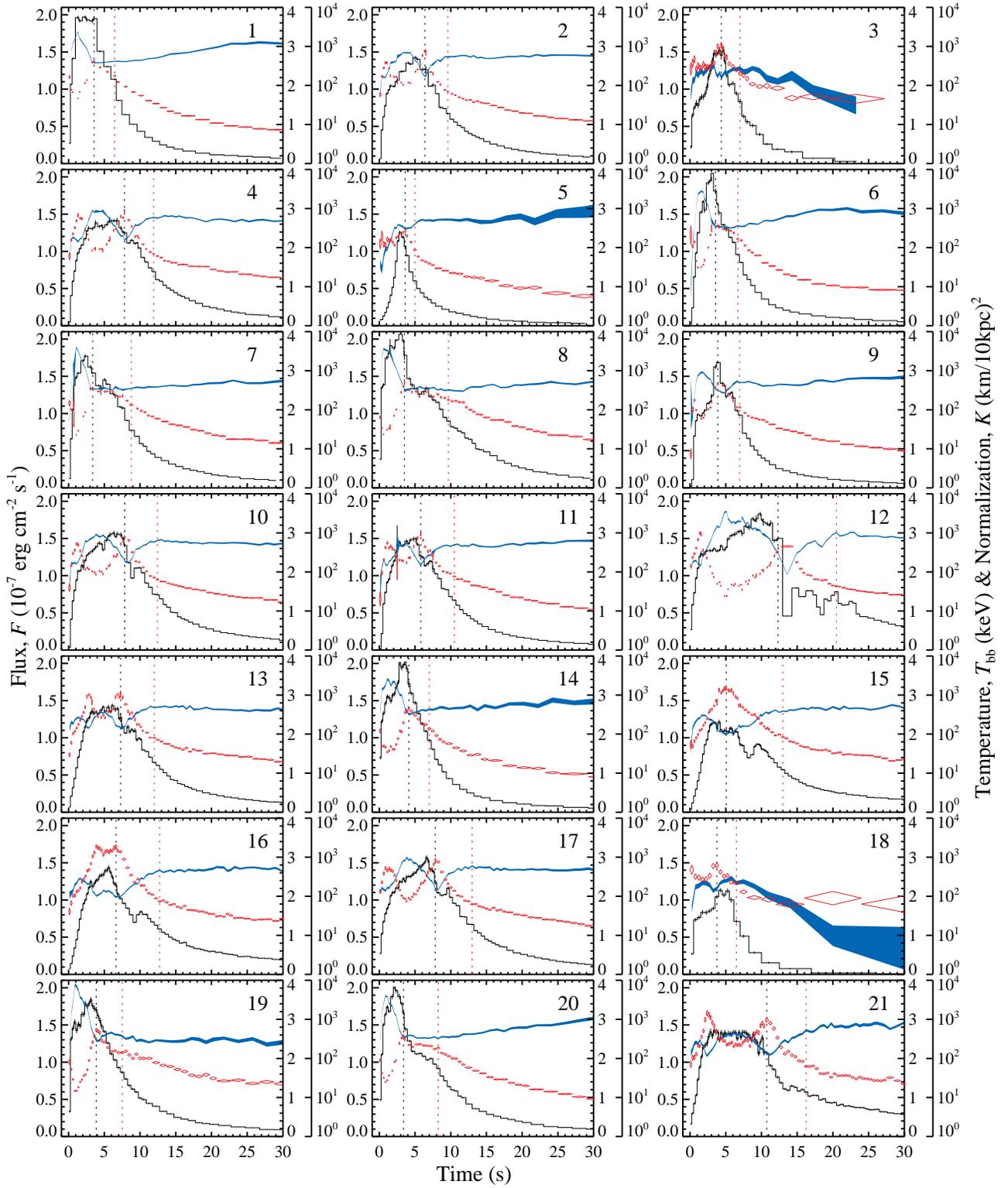}
\caption{Flux, temperature and black-body radius evolution during PRE bursts.
The black line in each panel shows the  bolometric flux (left-hand y-axis). 
The blue ribbon shows the 1$\sigma$ limits of the normalisation (outer right-hand y-axis).  
The red diamonds correspond to the 1$\sigma$ error box for black-body temperature (inner right-hand y-axis).  
Blue and red vertical dotted lines mark the touchdown and the time when the flux dropped to half the touchdown value, 
respectively. 
}	
\label{fig:lightcurve}
\end{figure*}

\begin{table*}
\caption{PRE X-ray bursts from 4U 1652--08.
\label{tab:bursts} }
\centering   
\begin{tabular}[c]{l c c c c c c c c c}
\hline
\#$^a$	& ID$^b$	& Start time$^c$	& $F_{\rm td,-9}$	$^d$ & $E_{\rm tot,-6}$$^e$	& $E_{\rm td,-6}$$^f$	& $\tau_{\rm td}$$^g$	& $F_{\rm per,-9}$$^h$ & $K_{\rm td/2}/K_{\rm td}$$^i$	& $S_{z}$$^j$ \\
\hline
1  & 30062-01-01-00 & 50899.58793 & 195$\pm$3 & 1.67$\pm$0.01 & 0.721$\pm$0.004 & 3.50$\pm$0.38  & 2.94$\pm$0.03  & 1.1$\pm$0.1 & 2.0 \\
2  & 30062-01-02-05 & 50914.27554 & 127$\pm$2 & 1.62$\pm$0.01 & 0.720$\pm$0.003 & 6.25$\pm$0.13  & 0.94$\pm$0.02  & 3.1$\pm$0.2 & 2.0 \\
3  & 50052-02-01-01 & 51612.03172 & 145$\pm$4 & 1.05$\pm$0.01 & 0.476$\pm$0.005 & 4.25$\pm$0.13  & 1.47$\pm$0.04  & 1.7$\pm$0.3 & 2.2 \\
4  & 50052-01-04-00 & 51614.07214 & 125$\pm$3 & 1.98$\pm$0.01 & 0.973$\pm$0.004 & 7.75$\pm$0.25  & 0.74$\pm$0.02  & 3.8$\pm$0.3 & 1.5 \\
5  & 70059-01-08-00 & 52499.40489 & 107$\pm$2 & 0.65$\pm$0.01 & 0.234$\pm$0.002 & 3.75$\pm$0.13  & 13.74$\pm$0.05 & 1.4$\pm$0.1 & 2.6 \\
6  & 70059-01-20-00 & 52524.10246 & 181$\pm$4 & 1.44$\pm$0.01 & 0.512$\pm$0.004 & 3.50$\pm$0.13  & 5.13$\pm$0.09  & 1.0$\pm$0.1 & 2.2 \\
7  & 70059-01-21-00 & 52526.16094 & 154$\pm$2 & 1.87$\pm$0.01 & 0.565$\pm$0.005 & 3.25$\pm$0.38  & 6.79$\pm$0.03  & 1.1$\pm$0.1 & 2.3 \\
8  & 70059-03-01-00 & 52529.18022 & 178$\pm$7 & 2.32$\pm$0.01 & 0.683$\pm$0.005 & 3.50$\pm$0.25  & 4.34$\pm$0.04  & 1.0$\pm$0.2 & 2.1 \\
9  & 70058-01-39-00 & 52536.31811 & 169$\pm$2 & 1.35$\pm$0.01 & 0.481$\pm$0.003 & 3.75$\pm$0.38  & 2.51$\pm$0.03  & 1.6$\pm$0.1 & 2.2 \\
10 & 70069-01-01-00 & 52542.50168 & 141$\pm$3 & 2.19$\pm$0.01 & 1.038$\pm$0.004 & 7.75$\pm$0.50  & 0.66$\pm$0.02  & 3.5$\pm$0.3 & 1.7 \\
11 & 70059-01-26-00 & 52546.90031 & 126$\pm$4 & 1.68$\pm$0.01 & 0.705$\pm$0.005 & 5.75$\pm$0.13  & 0.74$\pm$0.02  & 3.3$\pm$0.3 & 1.9 \\
12 & 80406-01-04-08 & 52727.18614 & 150$\pm$3 & 3.28$\pm$0.01 & 1.851$\pm$0.006 & 13.5$\pm$1.0 & 0.60$\pm$0.02  & 3.4$\pm$0.2 & 0.7 \\
13 & 90408-01-04-04 & 53104.40883 & 128$\pm$6 & 1.93$\pm$0.01 & 0.834$\pm$0.004 & 7.25$\pm$0.13  & 0.89$\pm$0.02  & 3.4$\pm$0.4 & 1.4 \\
14 & 93408-01-23-02 & 54434.97422 & 172$\pm$4 & 1.58$\pm$0.01 & 0.688$\pm$0.004 & 4.00$\pm$0.13  & 3.36$\pm$0.05  & 1.3$\pm$0.2 & 2.2 \\
15 & 93408-01-25-06 & 54452.11635 & 109$\pm$4 & 1.98$\pm$0.01 & 0.438$\pm$0.003 & 5.00$\pm$0.13  & 1.66$\pm$0.02  & 3.9$\pm$0.3 & 0.8 \\
16 & 93408-01-26-04 & 54461.03140 & 120$\pm$3 & 2.25$\pm$0.01 & 0.681$\pm$0.003 & 6.50$\pm$0.13  & 1.62$\pm$0.02  & 4.8$\pm$0.4 & 0.8 \\
17 & 93408-01-59-03 & 54692.07545 & 124$\pm$4 & 2.07$\pm$0.01 & 0.965$\pm$0.004 & 7.75$\pm$0.13  & 0.67$\pm$0.02  & 3.8$\pm$0.3 & 1.8 \\
18 & 94401-01-25-02 & 54997.68024 & 104$\pm$3 & 0.78$\pm$0.01 & 0.311$\pm$0.004 & 3.50$\pm$0.25  & 1.70$\pm$0.03  & 1.4$\pm$0.3 & 2.0 \\
19 & 95334-01-03-08 & 55270.22105 & 170$\pm$5 & 1.71$\pm$0.01 & 0.617$\pm$0.004 & 3.75$\pm$0.13  & 6.34$\pm$0.05  & 1.3$\pm$0.2 & 2.1 \\
20 & 96423-01-11-01 & 55725.15591 & 166$\pm$3 & 1.81$\pm$0.01 & 0.575$\pm$0.003 & 3.25$\pm$0.13  & 4.91$\pm$0.04  & 0.9$\pm$0.1 & 2.2 \\
21 & 96423-01-35-00 & 55890.37147 & 112$\pm$3 & 2.92$\pm$0.01 & 1.333$\pm$0.006 & 10.50$\pm$0.25 & 1.55$\pm$0.04  & 4.1$\pm$0.4 & 0.9 \\
\hline
\end{tabular}
\begin{flushleft}{ 
$^{a}$Burst number.\\
$^{b}$Observation ID during which the burst was observed. \\
$^{c}$Burst start time in MJD. \\
$^{d}$Touchdown flux in units of $10^{-9}$erg cm$^{-2}$ s$^{-1}$. \\
$^{e}$Burst fluence in units of $10^{-6}$ erg cm$^{-2}$. \\
$^{f}$Burst fluence from the burst start until the touchdown in units of  $10^{-6}$ erg cm$^{-2}$. \\
$^{g}$Time from the beginning of the burst to the touchdown (s). \\
$^{h}$Persistent flux level (in the interval 2.5--25~keV) prior to the burst  in units of  $10^{-9}$ erg cm$^{-2}$ s$^{-1}$.  \\
$^{i}$K-ratio, i.e. the ratio of the black-body normalisations  at 1/2 of the touchdown flux to that at the touchdown. \\
$^{j}$Value of $S_z$ of the object on the colour-colour diagram before the burst, $S_z\lesssim2$ corresponds to the low hard state, 
while $S_z\gtrsim2$ is in the high soft state.
  }\end{flushleft} 
\end{table*}

We analysed all publicly available \rxte\ data from 1995 December 30 through 2012 January 5.
During this time \rxte\ observed 56 bursts from \source\ of which 21 were recognised as PRE-bursts.
Time resolved spectra for the bursts were then extracted using an initial integration time of 0.25~s.
Then each time the count rate after the peak decreased by a factor of $\sqrt{2}$ the integration time was doubled. 
The exposure for each time bin was corrected for the dead-time 
following the approach recommended by the instrument team.
\footnote{\texttt{http://heasarc.gsfc.nasa.gov/docs/xte/recipes/pca\_deadtime.html}}
It resulted in a roughly 10--15 per cent increase in the peak flux, with the difference decreasing quickly as the observed flux drops. 
A spectrum extracted from a 16~s period prior to the burst  
was then subtracted as the background for each burst \citep[][and references therein]{Kuulkers02}. 
{We note  that variations in the persistent emission during the burst are possible, but  they are not significant 
in the cooling tail of the bursts \citep[see Fig.~6 in ][]{Worpel13}. 
We also checked that the difference in burst characteristics with and without background subtraction 
is negligible at least at high burst fluxes (at $F>0.2\,F_{\rm td}$).
}
These dead-time corrected spectra were then fitted with a black-body model multiplied by interstellar absorption.
For the  hydrogen column density, we adopt 
the value $N_{\rm H}=8.9\times 10^{21}$~cm$^{-2}$ obtained from the \textit{BeppoSAX} observations of \source\ \citep{Keek08}.
The best-fit parameters are the black-body (colour) temperature $T_{\rm bb}$ and 
the normalisation constant $K\equiv(R_{\rm bb} [{\rm km}]/D_{10})^2$, where $D_{10} \!=\! D/10$ kpc. 
The time-resolved spectral parameters  of analysed PRE-bursts are shown on Fig.~\ref{fig:lightcurve}.

\begin{figure}
\centering 
\includegraphics[width=6.5cm]{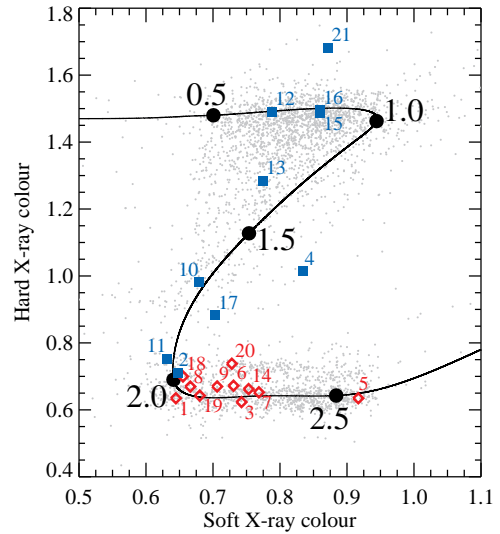}
\caption{Colour-colour diagram showing the $S_z$ parameter locus of the persistent spectrum before each PRE burst. 
The positions of the hard-state bursts are marked with blue squares and soft-state bursts with red diamonds.
The burst numbering follows Table~\ref{tab:bursts}. 
Grey dots show the positions of the object as determined from the data taken in 160~s long intervals.
}	
\label{fig:coco}
\end{figure}

 \begin{figure*}
\centering 
\includegraphics[width=12.5cm]{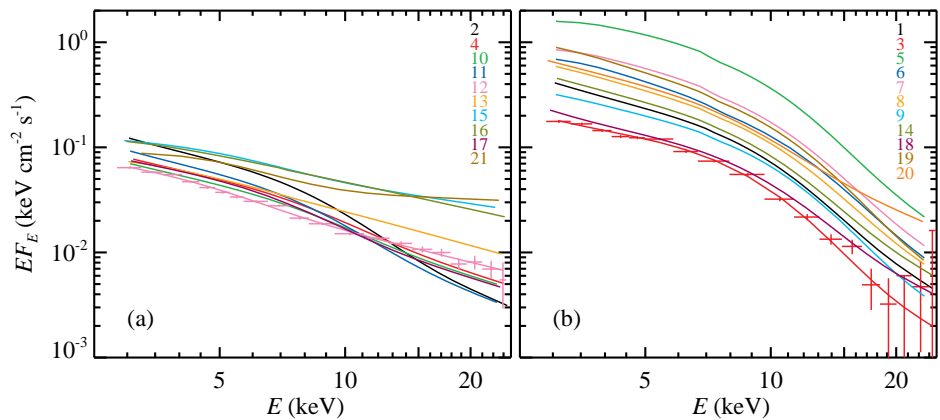}
\caption{Model spectra of the persistent emission before each PRE burst 
(a) for  $S_z\lesssim2$ (low hard state) and (b) for $S_z\gtrsim2$ (high soft state). 
The unfolded spectrum for one event is shown by crosses on each panel. 
The burst numbering follows Table~\ref{tab:bursts} and colours are used only for easier identification. 
}	
\label{fig:preburst_spec}
\end{figure*}

These bursts show typical characteristics of photospheric radius expansion: 
fast rise of the flux at the beginning and peak in the normalisation after a few seconds.
The temperature evolution of these bursts also shows the characteristic double-peaked structure,
arising from the cooling of the photosphere when it expands and the subsequent heating when it collapses back towards the surface. 
Because of this expansion, we can assume that the flux has reached the Eddington limit.
The moment when the temperature reaches its second peak and normalisation its minimum is defined to be the touchdown
\citep[but see][]{SLB10}, where the atmosphere has collapsed back to the NS surface 
(due to the data gaps during burst 12, the touchdown was defined there to be just before the gap).
This also marks the beginning of the cooling phase where normalisation rises to a nearly constant level while the flux and the temperature 
continue to decrease for the rest of the burst.

From the flux evolution, we have determined different characteristics of the bursts such as 
the peak flux, the total burst fluence, the burst fluence until the touchdown, and 
the time from the beginning of the burst to the touchdown. 
For every PRE-burst, we have obtained the dead-time corrected spectrum of the persistent emission 
using 160~s long interval just before the burst.  
These spectra were then fitted with a model consisting of a black body ({\sc bbodyrad}), 
comptonization ({\sc comptt}) component \citep{Titarchuk94} and an iron line with the energy fixed at 6.4 keV, 
attenuated by interstellar absorption ({\sc phabs}). 
The observed source flux over the energy range 2.5--25~keV was estimated using  the {\sc cflux}-model of {\sc xspec}.
In order to characterise the persistent spectrum before the bursts, 
we also computed hard and soft X-ray colours as the ratio of fluxes in the 
(8.6--18.0)/(5.0--8.6)~keV and (3.6--5.0)/(2.2--3.6)~keV energy bands (see Fig.~\ref{fig:coco}). 
From these colours we were able to define the $S_z$ coordinate locus using a similar method as in \cite{GMH08}.
The $S_{z}$ is  thought to be related to mass accretion rate but the exact dependence is not known \citep{vdK95}. 
The spectra of the persistent emission are presented in Fig.~\ref{fig:preburst_spec}. 
We have separated them into two groups depending  the value of $S_z$ and the shape of the spectrum: 
the left panel shows the spectra with $S_z\lesssim2$ (hard state), while the right panel is for $S_z\gtrsim2$ (soft state).
All obtained parameters and associated 1$\sigma$ errors are listed in Table \ref{tab:bursts}.

In addition to the PRE bursts observed by \rxte, 
we have used the time-resolved X-ray spectral fits of two exceptional bursts observed 
by \textit{EXOSAT}/ME in 1984 July 5 and 1986 March 12
during the hard state and a very low persistent flux  \citep{Gottwald87,Penninx89}.

\section{Results}

\subsection{Burst spectral evolution and its relation to the persistent emission} 
\label{sec:preburst}

\begin{figure*}
\centering 
\includegraphics[width=16.5cm]{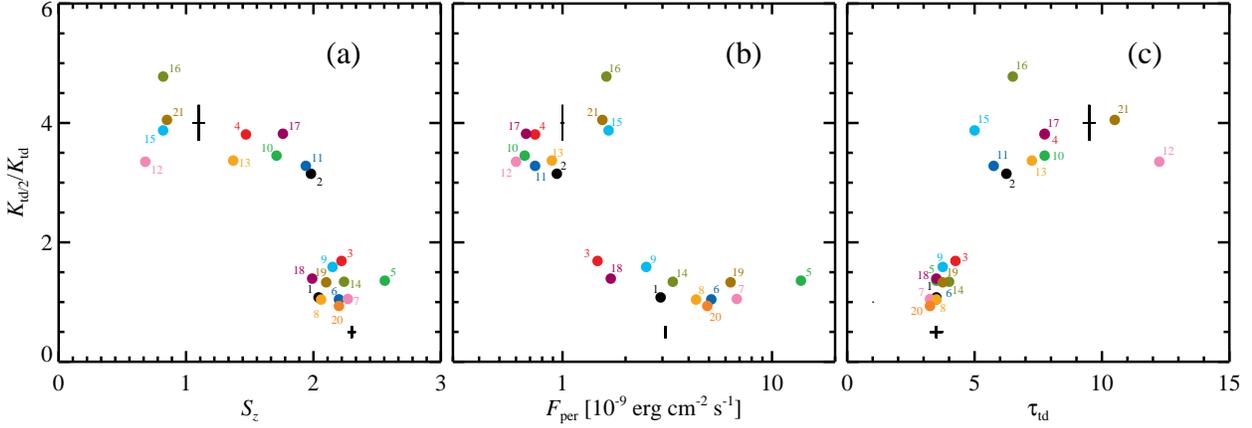}
\caption{The K-ratio (i.e. the ratio of the black-body normalisation at 1/2 of the touchdown flux $K_{\rm td/2}$ 
to that at the touchdown $K_{\rm td}$) as a function of the (a) $S_z$ value, 
(b) persistent flux $F_{\rm per}$ (in 2.5--25 keV band) before the burst, and 
(c) time between the burst start and the touchdown $\tau_{\rm td}$. 
Typical error bars are shown by black crosses. 
Bursts are numbered according to the order in Table~\ref{tab:bursts} and 
identified by the same colours as in Fig.~\ref{fig:preburst_spec}. 
}	
\label{fig:k2k1}
\end{figure*}

\begin{figure*}
\centering 
\includegraphics[width=7.cm]{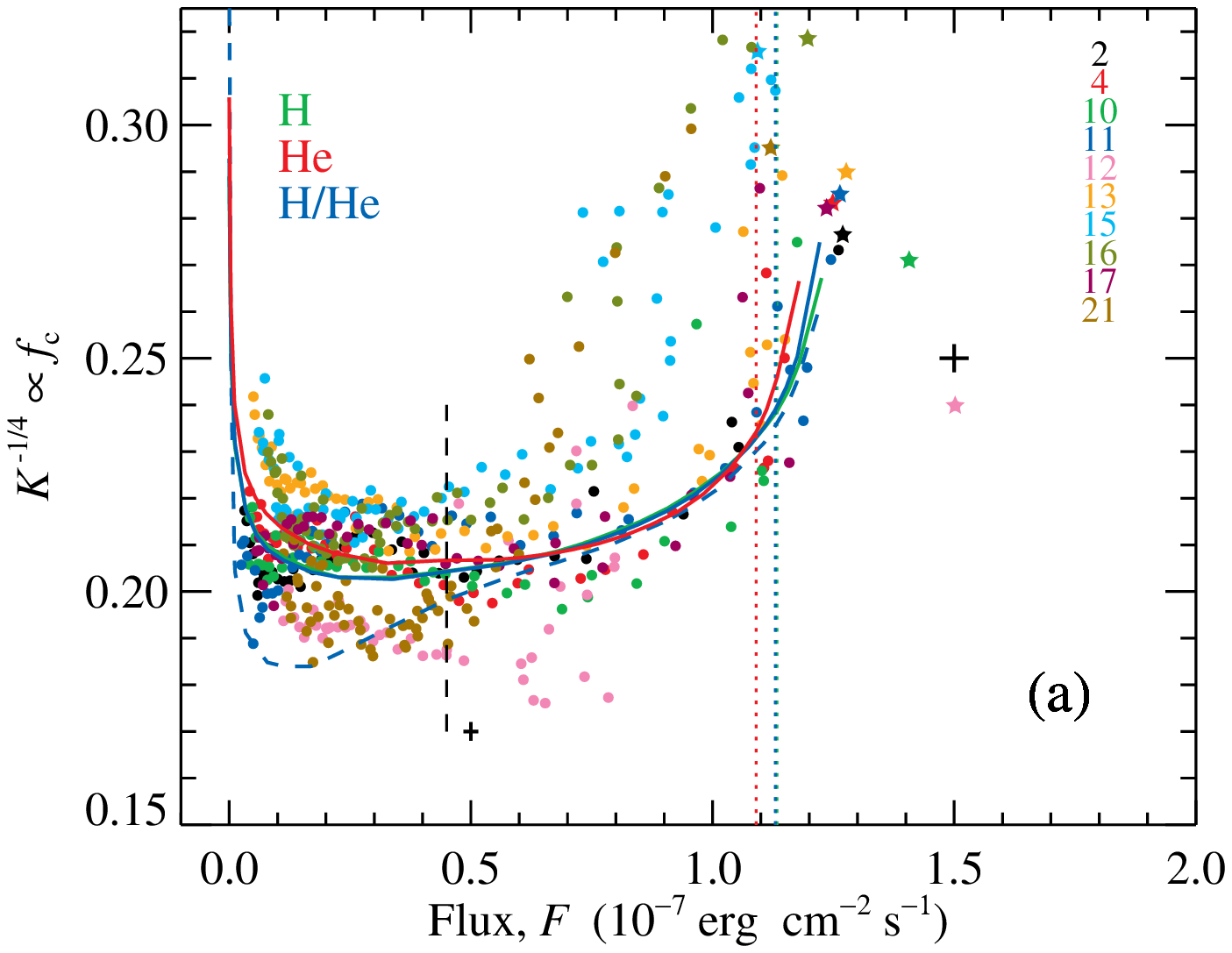}
\hspace{1cm}
\includegraphics[width=7.cm]{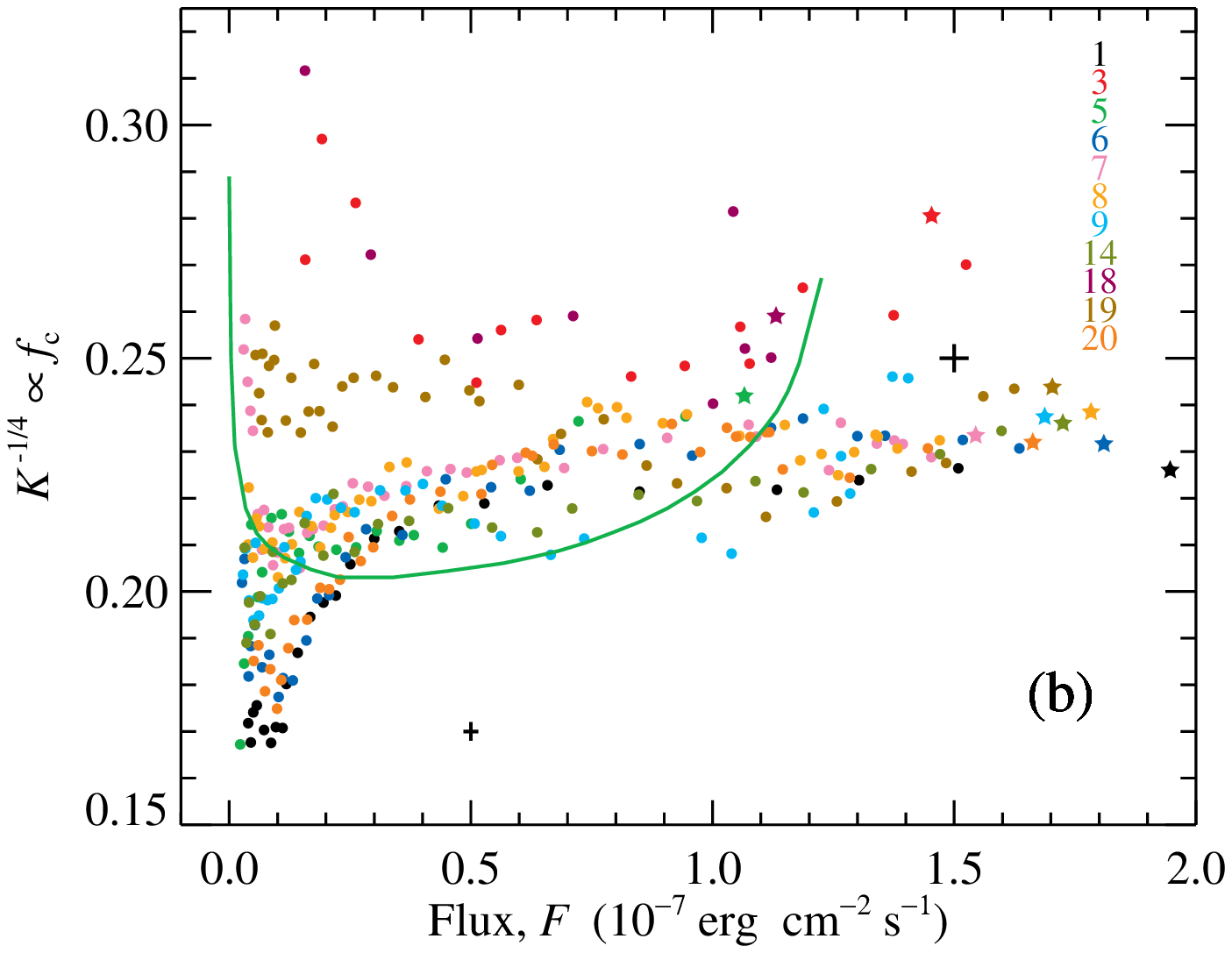}
\caption{
Evolution of $K^{-1/4}$ with flux during the cooling tail  
of the (a) hard-state (low accretion rate) and (b) soft-state (high accretion rate) bursts.  
The cooling tracks for different bursts are shown from the touchdown (marked by stars) 
to the end of the burst.
The best-fitting theoretical models $A\,f_{\rm c}(F/\fedd)$ 
for $\log g=14$ \citep{SPW12} to the combined data from the hard-state bursts 2, 4, 10, 11, 13 and 17 
are shown by solid curves (pure helium -- red, pure hydrogen -- green,  
and solar mixture of H and He with sub-solar metal abundances of  $Z \!=\! 0.01 Z_{\odot}$ -- blue, which nearly coincides with the green curve). 
The vertical dotted lines indicate the position of the Eddington flux $F=\fedd$.  
The dashed line marks the position of the minimum flux $F_{\min}=F_{\rm td}/e$ used in most of the fits. 
The black crosses indicate the typical error bars. 
The model for solar composition ($Z \!=\! Z_{\odot}$) shown by the dotted blue curve 
does not describe well the data at fluxes below half of the touchdown flux.  
For the soft-state bursts,  the models do not fit the data well and only 
the hydrogen model from panel (a) is shown to guide the eye.  
}
\label{fig:K14_f}
\end{figure*}

Spectral  evolution of the bursts during the cooling tail contains the information about NS compactness. 
It is this part of the burst that should be compared to theoretical models of NS atmospheres 
in order to constrain the mass and radius using the cooling tail method 
\citep[see Appendix~\ref{app:A} and][ for a full description of the method]{SPW11}. 
Important assumptions of the method are that  during the cooling tail  
there are no eclipses of the NS and that the burst spectrum is formed in a passively cooling NS atmosphere 
(i.e. not influenced by the accretion flow).  
In that case, the theory predicts that $K^{-1/4}$ is proportional to the colour-correction factor $f_{\rm c}$,
which falls from  a value exceeding 1.8 at the touchdown when $L  \! \approx \!  L_{\rm Edd}$ 
to  $f_{\rm c}  \lesssim 1.5$ at $L  \! \approx \!  0.5 L_{\rm Edd}$ \citep{SPW11,SPW12}. 
If the Eddington luminosity is reached near the touchdown, the  ratio of observed black body normalisations 
at half the touchdown flux to that at the touchdown $K_{\rm td/2} /K_{\rm td}$ (which we will call the K-ratio)  should exceed 2 
(just because $f_{\rm c}(L_{\rm Edd})/f_{\rm c}(L_{\rm Edd}/2)\gtrsim 1.2$). 
The data, however, show a clear dependency of the K-ratio on the value of $S_z$  (see Fig.~\ref{fig:k2k1}a),
with a number of bursts having $K_{\rm td/2} /K_{\rm td}<2$. 
It also depends on other model-independent parameters (see Table~\ref{tab:bursts}) derived from the bursts (Figs~\ref{fig:k2k1}b,c),
such as the persistent flux prior to the bursts, and 
the duration of the phase prior to the touchdown. 
The bursts can be now separated into two distinct groups.
The first group   (bursts 2, 4, 10--13, 15--17 and 21)  occur at $S_z\lesssim2$ at low persistent fluxes and have 
 $K_{\rm td/2} /K_{\rm td}>2$ consistent with those predicted by the atmosphere models.
The second group of bursts (1, 3, 5--9, 14, 18--20) happening at $S_z\gtrsim2$, at higher persistent fluxes, 
on the other hand, has $K_{\rm td/2} /K_{\rm td}<2$, inconsistent with theoretical predictions. 
Thus, the bursts can be cleanly separated into groups either by their K-ratio, or 
based on the shape of the spectrum of the persistent emission prior to the burst (Fig.~\ref{fig:preburst_spec}), 
using, e.g., the source position on the colour-colour diagram (see Fig.~\ref{fig:coco}), or 
duration of the super-Eddington phase $\tau_{\rm td}$, or the persistent flux  (see Fig.~\ref{fig:k2k1}).

The bursts from the first group occur at low accretion rate (with $F_{\rm per}\lesssim0.015 F_{\rm Edd}$), 
when the object is in the hard state. Here the persistent 
spectra are closer to a power-law, produced most probably either in the hot inner flow of the accretion disc 
or the optically thin boundary layer \citep{KW91,PS01}. In this situation, the evolution of $K^{-1/4}$ with flux 
during the bursts follows the theoretical models with very little metals \citep{SPW11,SPW12} 
down to rather low  luminosities (see Fig.~\ref{fig:K14_f}(a)). 
The exception are two bursts: 12 and 21 (see pink and brown points in Fig.~\ref{fig:K14_f}(a)). 
Both show clear drop in $K^{-1/4}$ value at fluxes 0.3--0.5 of the touchdown flux. 
Interestingly, both bursts demonstrate two  times longer ($\sim$10~s vs $\sim$5~s) 
super-Eddington phase until the touchdown (see Table~\ref{tab:bursts}). 
We can speculate that in these bursts a lot of accreted material was blown away during this phase 
exposing the material rich in heavy elements (see \citealt{WBS06}), 
resulting in strong edges in the observed X-ray band and reduction of the colour-correction factor \citep{SPW11,SPW12}. 
 
\begin{figure*}
\centering 
\includegraphics[width=11cm]{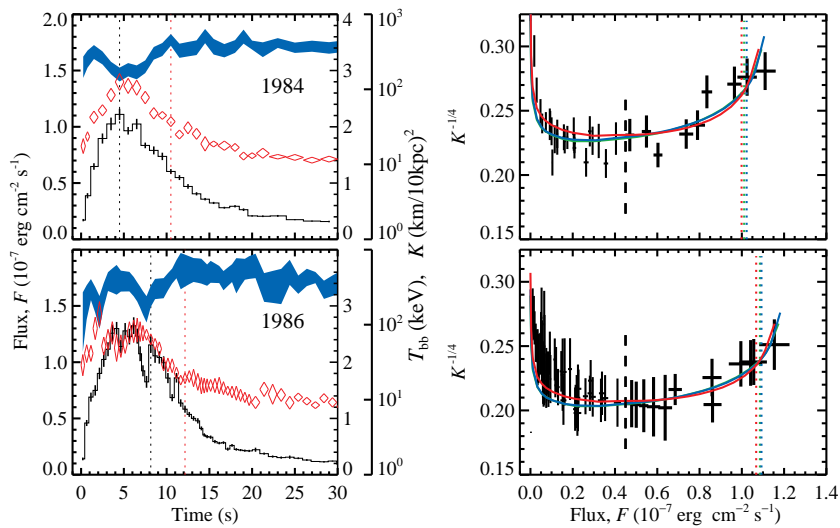}
\caption{Left: Time evolution of flux, temperature and black-body normalisation of the 
hard-state X-ray bursts observed by \textit{EXOSAT} in 1984 and 1986 \citep{Gottwald87,Penninx89}. 
Right:  Same as Fig.~\ref{fig:K14_f}, but for the \textit{EXOSAT} bursts. }	
\label{fig:exosat}
\end{figure*}

The bursts from the second group occur at high persistent fluxes (with $F_{\rm per}\gtrsim0.015 F_{\rm Edd}$), 
when the source was in the soft state with 
the spectrum dominated by the black-body-like radiation in the soft X-rays, probably 
coming from the boundary layer and the accretion disc \citep{GRM03,RG06,SulP06,RSP13}. 
These bursts have almost constant normalisation $K$ over a substantial range of luminosities, with a deviation seen only at 
fluxes below 20 per cent of the touchdown flux (see Fig.~\ref{fig:K14_f}(b)).  
This evolution is not consistent with NS atmosphere models and therefore these bursts cannot be used for further interpretation with 
the aim to measure NS parameters. 
It is clear that some of the assumptions that the models are based on are not valid for those bursts.  
The most obvious explanation is that at higher accretion rates, the accreting matter 
influences the atmosphere by forcing the upper layer  to rotate \citep{IS99,SulP06} and probably by heating them. 
This then affects the emerging spectrum so that the assumption of the passively cooling atmosphere is not valid anymore. 
Thus, it is clear that only the hard-state bursts, that show the predicted evolution, should be used to constrain NS mass and radius.

\subsection{NS mass and radius in \source\ from the hard-state bursts}
\label{sec:hard}

\begin{table*}
\centering   
\caption{Parameters of the fits of the 
$K^{-1/4}-F$ dependency with the NS atmosphere models for various chemical compositions and $\log g \!=\!  14.0$. 
\label{tab:results} 
}
\begin{minipage}{120mm}
\begin{tabular}[c]{l c c c  c }
\hline
\# & Composition & $F_{\rm{Edd}}$ & $A$ & $T_{\rm Edd,\infty}$ \\ 
 &   & ($10^{-7}$ erg cm$^{-2}$ s$^{-1}$) & ((km/10~kpc)$^{-1/2}$)   &  ($10^7$~K) \\ 
\hline 
\multicolumn{5}{c}{Individual bursts} \\
2 & H & 1.16$\pm$0.05 & 0.137$\pm$0.003 & 1.62$\pm$0.03 \\
 & H/He & 1.17$\pm$0.03 & 0.138$\pm$0.003 & 1.64$\pm$0.03 \\
 & He & 1.05$\pm$0.05 & 0.143$\pm$0.002 & 1.65$\pm$0.03 \\
4 & H & 1.10$\pm$0.03 & 0.133$\pm$0.002 & 1.55$\pm$0.03 \\
 & H/He & 1.10$\pm$0.04 & 0.134$\pm$0.003 & 1.57$\pm$0.03 \\
 & He & 1.08$\pm$0.04 & 0.140$\pm$0.003 & 1.62$\pm$0.03 \\
10 & H & 1.15$\pm$0.03 & 0.132$\pm$0.003 & 1.56$\pm$0.03 \\
 & H/He & 1.14$\pm$0.03 & 0.133$\pm$0.003 & 1.57$\pm$0.03 \\
 & He & 1.11$\pm$0.03 & 0.139$\pm$0.003 & 1.63$\pm$0.03 \\
11 & H & 1.17$\pm$0.04 & 0.140$\pm$0.003 & 1.65$\pm$0.03 \\
 & H/He & 1.17$\pm$0.04 & 0.141$\pm$0.003 & 1.67$\pm$0.03 \\
 & He & 1.16$\pm$0.04 & 0.148$\pm$0.003 & 1.75$\pm$0.03 \\
13 & H & 1.08$\pm$0.04 & 0.140$\pm$0.003 & 1.62$\pm$0.03 \\
 & H/He & 1.09$\pm$0.04 & 0.141$\pm$0.003 & 1.64$\pm$0.04 \\
 & He & 1.07$\pm$0.04 & 0.148$\pm$0.003 & 1.71$\pm$0.03 \\
15 & H & 0.99$\pm$0.04 & 0.149$\pm$0.003 & 1.69$\pm$0.04 \\
 & H/He & 1.00$\pm$0.04 & 0.150$\pm$0.003 & 1.71$\pm$0.04 \\
 & He & 0.97$\pm$0.04 & 0.156$\pm$0.003 & 1.76$\pm$0.04 \\
16 & H & 0.85$\pm$0.03 & 0.143$\pm$0.003 & 1.56$\pm$0.04 \\
 & H/He & 0.84$\pm$0.03 & 0.144$\pm$0.003 & 1.57$\pm$0.04 \\
 & He & 0.83$\pm$0.03 & 0.151$\pm$0.004 & 1.64$\pm$0.04 \\
17 & H & 1.14$\pm$0.04 & 0.138$\pm$0.003 & 1.62$\pm$0.04 \\
 & H/He & 1.13$\pm$0.05 & 0.139$\pm$0.003 & 1.63$\pm$0.04 \\
 & He & 1.06$\pm$0.03 & 0.144$\pm$0.004 & 1.66$\pm$0.04 \\
Exo1$^a$ & H & 1.01$\pm$0.05 & 0.153$\pm$0.006 & 1.74$\pm$0.07 \\
 & H/He & 1.02$\pm$0.05 & 0.155$\pm$0.006 & 1.77$\pm$0.07 \\
 & He & 1.00$\pm$0.06 & 0.161$\pm$0.005 & 1.84$\pm$0.06 \\
Exo2$^b$ & H & 1.09$\pm$0.05 & 0.137$\pm$0.006 & 1.60$\pm$0.07 \\
 & H/He & 1.09$\pm$0.06 & 0.139$\pm$0.006 & 1.62$\pm$0.07 \\
 & He & 1.07$\pm$0.05 & 0.145$\pm$0.006 & 1.67$\pm$0.07 \\
\multicolumn{5}{c}{Combined bursts 2, 4, 10, 11, 13 and 17} \\
All$^c$ & H & 1.13$\pm$0.05 & 0.137$\pm$0.003 & 1.61$\pm$0.04 \\
 & H/He & 1.13$\pm$0.05 & 0.138$\pm$0.003 & 1.62$\pm$0.04 \\
 & He & 1.09$\pm$0.06 & 0.144$\pm$0.002 & 1.68$\pm$0.03 \\
All$^d$ & H & 1.13$\pm$0.06 & 0.137$\pm$0.004 & 1.61$\pm$0.05 \\
 & H/He & 1.11$\pm$0.06 & 0.138$\pm$0.003 & 1.61$\pm$0.04 \\
 & He & 1.11$\pm$0.05 & 0.145$\pm$0.003 & 1.70$\pm$0.04 \\
All$^e$ & H & 1.17$\pm$0.05 & 0.140$\pm$0.001 & 1.66$\pm$0.02 \\
 & H/He & 1.17$\pm$0.04 & 0.141$\pm$0.001 & 1.68$\pm$0.02 \\
 & He & 1.11$\pm$0.05 & 0.145$\pm$0.001 & 1.69$\pm$0.02 \\
 \hline
\end{tabular}
\begin{flushleft}{ 
Note: Errors correspond to the 90\% confidence level.\\ 
$^a$Burst observed by \textit{EXOSAT}   in 1984. \\
$^b$Burst observed by \textit{EXOSAT}   in 1986. \\
$^{c}$Best-fitting parameters for the combined data for bursts  2, 4, 10, 11, 13 and 17 
with the lower limit on the flux $F_{\min}=F_{\rm td}/e$. \\
$^{d}$Same as case $^c$, but for $F_{\min}=0.5~F_{\rm td}$. \\
$^{e}$Same as case $^c$, but for $F_{\min}=0.1~F_{\rm td}$.  
}\end{flushleft} 
\end{minipage}
\end{table*}

\begin{figure}
\centering
\includegraphics[width=7.0cm]{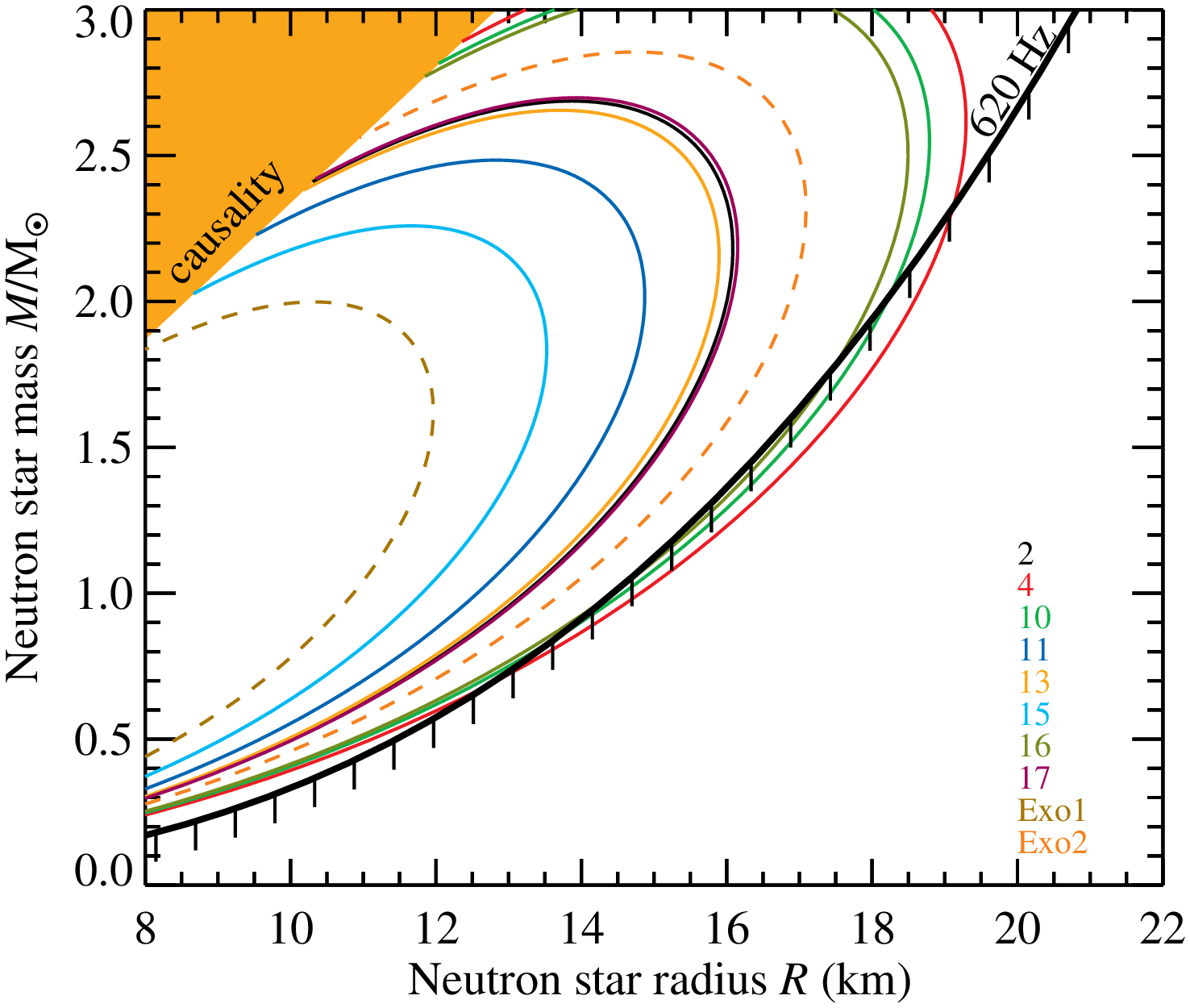}
\caption{Mass-radius constraints coming from individual hard-state bursts of \source\  
based only on the measured value of  $T_{\rm Edd,\infty}$ assuming hydrogen atmosphere. 
The upper-left region is excluded by constraints from the causality requirements \citep{HPY07,LP07}. 
The  constraints for the \textit{EXOSAT}  bursts are shown by the dashed curves. 
The lower-right region marked by black curve with downward ticks  lies below the 
mass-shedding limit  \citep{LP07} for the observed rotational frequency of 620~Hz.
}
\label{fig:mr_hard_individ}
\end{figure}

First, we consider the constraints that can be obtained from individual bursts. 
Bursts 12 and 21  showing significantly different evolution were not included in further studies. 
We use the cooling tail method \citep[see Appendix~\ref{app:A} and][]{SPRW11} that requires that the 
evolution of the black-body normalisation with flux $F$ after the touchdown 
(when the  atmosphere radius is assumed to coincide with the NS radius)
is to be described by the theoretical models of the evolution of the colour-correction factor. 
The dependence $K^{-1/4}$ on the observed flux $F$ is fitted by theoretical curves $A\,f_{\rm c}(F/\fedd)$ 
obtained from the most recent hot NS atmosphere models that account 
for Klein-Nishina reduction of the electron scattering opacity \citep{SPW12} for three chemical compositions. 

Because the data have errors in both directions, there are outliers and 
the distribution of points around the best-fitting curve 
does not follow a Gaussian, the $\chi^2$-statistics is not appropriate. 
Instead, we use a robust maximum likelihood estimator \citep{Press_numrec} 
and minimise the merit function 
\begin{equation}
L = \sum_i  \ln \left(1+\frac{z_i^2}{2} \right).
\end{equation} 
This introduces a Lorentzian weighting function into our maximum likelihood estimator that then removes the contribution from the most deviant outlier points 
but acts naturally with points that are close by. 
Here $z_{i}$ is the normalised minimum distance of the $i$th data point from the model curve $\hat{y}(\hat{x})$: 
\begin{equation}
z_{i}^2 = \frac{ \displaystyle \left( \frac{x_{i} - \hat{x} }{  \bar{\sigma}_x } \right)^2 + \left( \frac{ y_{i} - \hat{y} }{ \bar{\sigma}_y } \right)^2 } 
{ \displaystyle \left(\frac{\sigma_{x_i}}{\bar{\sigma}_x} \right)^2 + \left(\frac{\sigma_{y_i}}{ \bar{\sigma}_y} \right)^2 },
\end{equation}
where $(x_i, y_i)$ are the coordinates for the $i$th data point (substitute $F$ for $x$ and $K^{-1/4}$ for $y$) , 
$(\sigma_{x_i}, \sigma_{y_i})$ are the errors, and $\bar{\sigma}_x$ and $\bar{\sigma}_y$ are the 
mean errors in the $x$-  and $y$-direction over all data points. 
The uncertainties in the best-fitting parameters are obtained with the bootstrap method. 

The free parameters for the fits are  the quantity $A \!=\! (R_{\infty}~[{\rm km}] / D_{10})^{-1/2}$, which 
is related to the observed NS radius at the infinity $R_{\infty}=R(1+z)$ (here $R$ is the circumferential 
NS radius and $z$ is the surface redshift) and the Eddington flux $\fedd$. 
We choose the atmosphere models with $\log g \!=\!  14.0$, 
because the results are rather insensitive to its choice (see below). 
Compositions considered are pure hydrogen (H), pure helium (He) and solar composition of H 
and He with sub-solar metal abundance of $Z \!=\! 0.01 Z_{\odot}$ (H/He). 
It seems that $Z \!<\! 0.1 Z_{\odot}$ in the surface layers, because in the opposite case 
the atmosphere model predicts a drop in $f_{\rm c}$ (and correspondingly in  $K^{-1/4}$) 
at $F\sim 0.3 \fedd$ \citep{SPW11,SPW12}, which is not observed. 
The low metal abundance might be caused by chemical stratification. 
For example, in mostly hydrogen atmosphere with the  surface gravity $\log g=14$ 
at $T\sim10^7$~K  and density at the photosphere of $\rho\approx0.2$~g~cm$^{-3}$, 
the iron settling time-scale is $<10^{-3}$~s \citep[see equation (9) in][]{BBC02}. 
Alternatively, the absence of that drop can result from the temperature inversion because of accretion. 

We also consider two bursts observed by \textit{EXOSAT}  during a very low hard state \citep{Gottwald87,Penninx89},
in spite of the fact that they do not seem to be PRE.
We note that the cooling tail method allows to determine the Eddington flux even for bursts not reaching 
the Eddington limit, because of the curvature in the   $K^{-1/4}$ vs. $F$ dependence. 
For the two \textit{EXOSAT}  bursts, these dependences are well described by theoretical atmosphere models
and are similar to the PRE hard-state bursts observed by \rxte\ (see Fig.~\ref{fig:exosat}).

 \begin{figure*}
\centering
\includegraphics[width=7.0cm]{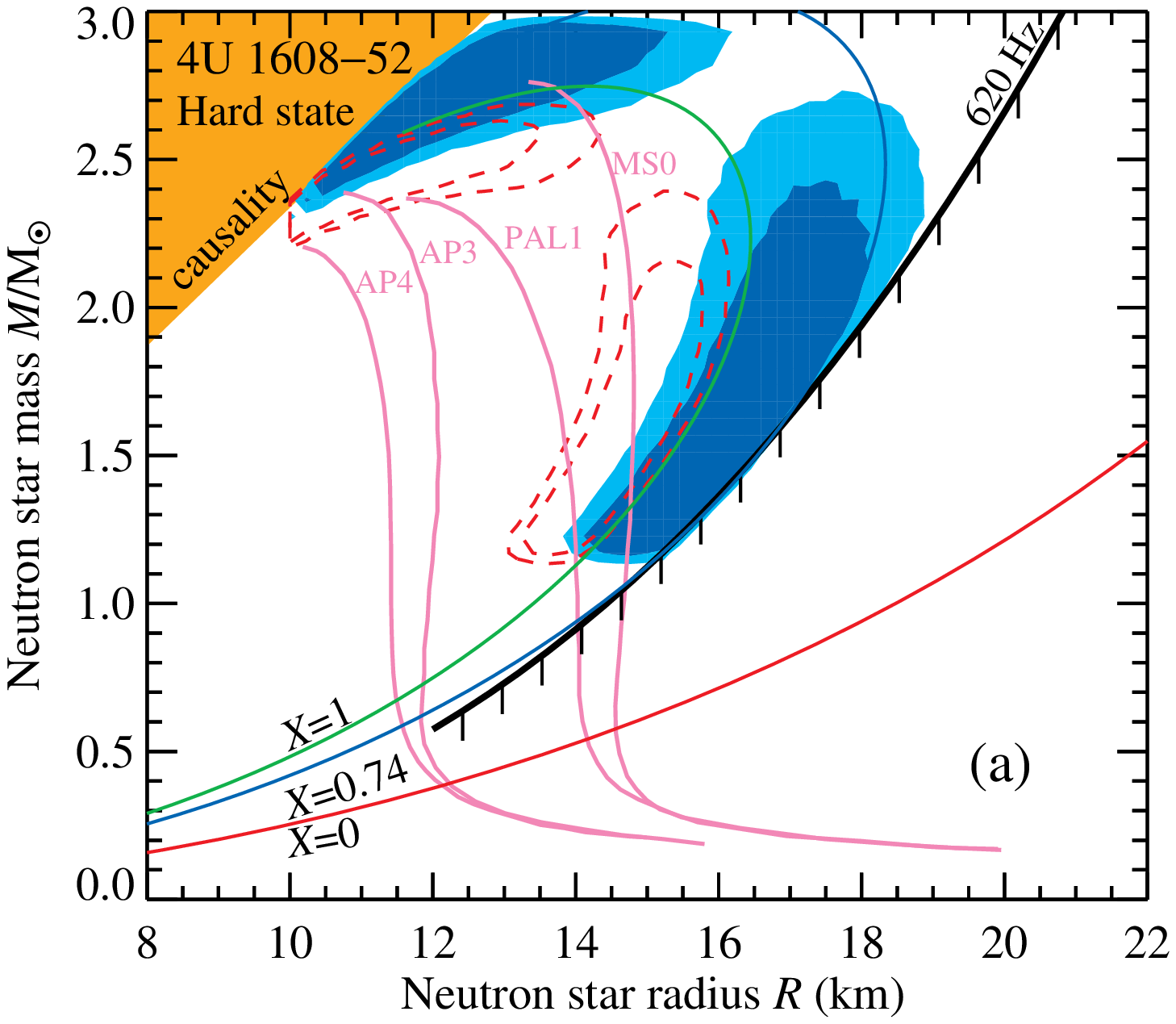}\hspace{1cm}
\includegraphics[height=7.cm,angle=90]{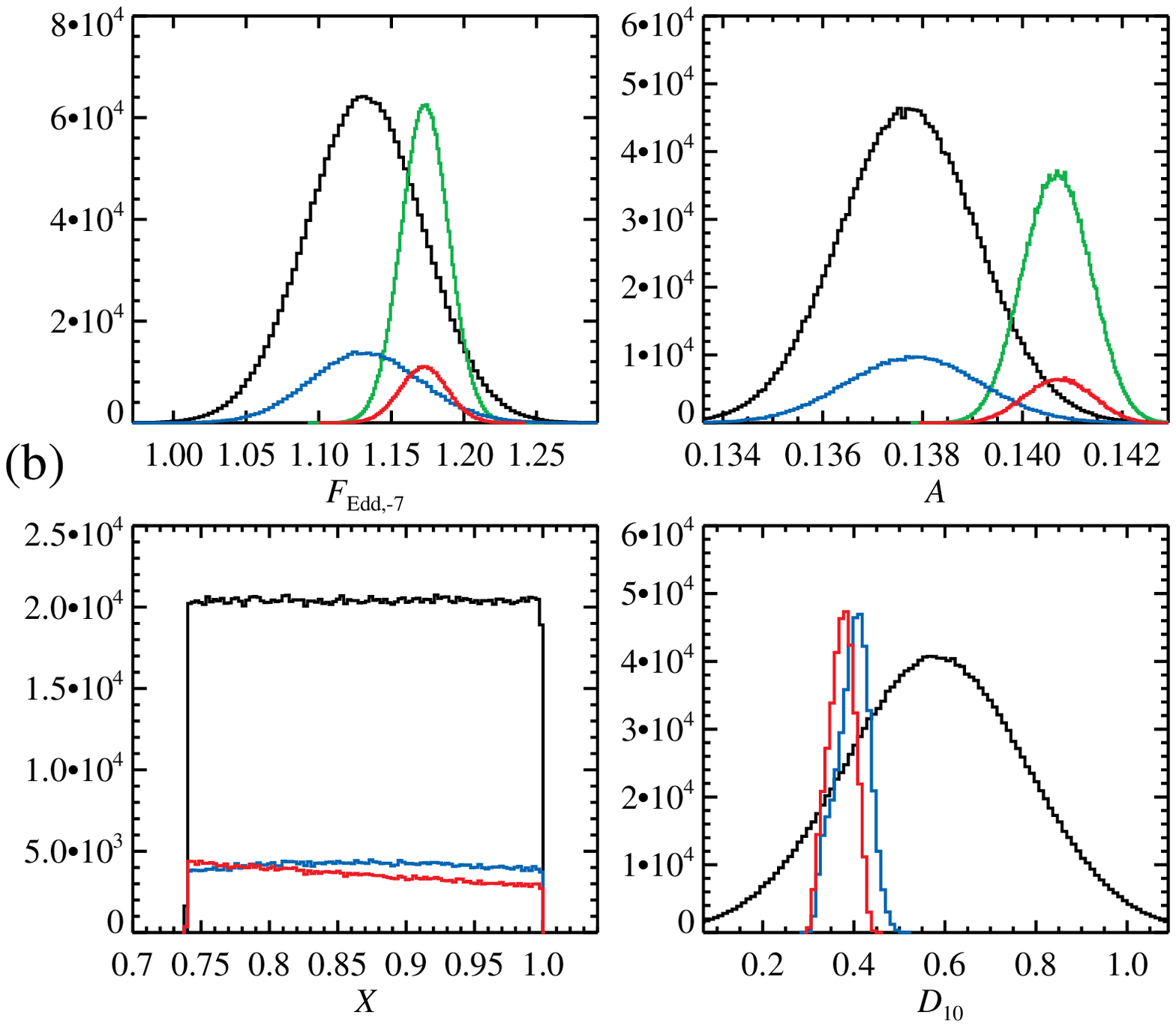} 
\caption{(a) Mass-radius constraints from the hard-state bursts of \source\ for a distance of $5.8\pm2.0$~kpc 
assuming $1.2\msun<M<3\msun$ and $0.74<X<1$.  
The dark and light blue contours correspond to 90 and 68 per cent confidence limits for $F_{\min}=F_{\rm td}/e$. 
For $X\lesssim0.7$ solutions lie below the mass shedding limit  for a rotational frequency of 620~Hz 
marked by the black curve with downward ticks. 
The dashed red contours are similar constraints for $F_{\min}=0.1\,F_{\rm td}$. 
Solid green, blue and red curves correspond to the best-fitting $T_{\rm Edd,\infty}$ for the combined bursts 
(with $F_{\min}=F_{\rm td}/e$), assuming H, solar H/He ratio and He composition, respectively. 
The NS mass-radius relations for several equations of state of cold dense matter 
that do not contradict the existence of 2$\msun$ pulsars  \citep{Demorest10,Antoniadis13} are shown by pink curves. 
(b) Corresponding distributions of parameters from the analysis of the hard-state bursts. 
The black and green histograms are the prior parameter distributions,  
and the  blue and red histograms are the posterior distributions of parameters that 
give a physical solution (see Appendix~\ref{app:A}) 
for $F_{\min}=F_{\rm td}/e$ and $F_{\min}=0.1\,F_{\rm td}$,  respectively. 
For some parameters  there are two solutions, therefore 
the  posterior distributions can  exceed the prior one.
}	
\label{fig:mr_hard}
\end{figure*}

The best-fitting values for $A$ and $\fedd$ for the hard-state bursts 
(using the data from touchdown to $F_{\min}=F_{\rm td}/e$) are presented in Table~\ref{tab:results}.
Note that the values of $\fedd$ are typically smaller by 10--15 per cent than the 
touchdown flux, mostly because of the difference between the actual electron scattering 
opacity and the Thomson one for which the Eddington flux is defined \citep{SPW12}. 
We also present in Table~\ref{tab:results} the values of the Eddington temperature $T_{\rm Edd,\infty}$.  
This is an observable that allows to get a  distance-independent constraint on the NS mass-radius relation (see Appendix~\ref{app:A}).   
We see that most of the bursts give very similar  results, except bursts 15, 16 and the \textit{EXOSAT} burst from 1984 (Exo1), 
which show significantly smaller $\fedd$ and a larger $A$ (i.e. much smaller black-body normalisation $K$), 
which could be a result of a confined burning. 
They also might not be genuine PRE bursts as the temperature profile does not show a clear two-peak structure
seen in other bursts.      
For illustration,  we present in Fig.~\ref{fig:mr_hard_individ}  
the mass-radius constraints obtained from the measured values of $T_{\rm Edd,\infty}$  
not accounting for errors and constraints on the distance assuming hydrogen atmosphere. 
We see that all bursts (except bursts 15 and Exo1 showing lower emitting area) 
give minimum NS radius of 13 km at $1.2\msun<M<2.4\msun$. 
For other atmosphere compositions, the radii are even larger. 
We note that pure He composition can be safely rejected, because it 
predicts a mass much below the mass-shedding limit for a star rotating at 620 Hz \citep{MC02,GMH08}. 
Solar  H/He ratio is barely consistent with this constraint. 
The fact that only hydrogen-rich models give solutions above the mass-shedding limit is 
consistent with the orbital period of the system \citep[$\sim$0.5 days, see][]{Wachter02} and inferred companion.

We have selected \rxte\ bursts 2, 4, 10, 11, 13 and 17  that show consistent results and 
larger emitting area (increasing the chance that burning is happening over the whole NS surface, which is also visible) 
to construct a combined cooling track.\footnote{We note here that the hard-state bursts  excluded from this analysis 12, 15, 16 and 21, 
all occur at a low state at $S_z<1$.  }
As in the case of individual bursts, we use the data down to $F_{\min}=F_{\rm td}/e$. 
The best-fitting atmosphere models  for all considered chemical compositions  
are shown by solid curves in Fig.~\ref{fig:K14_f}(a) and the parameters are presented in Table~\ref{tab:results}.
The parameters lie very close to the  mean values obtained from individual bursts. 
The black-body normalisation in the cooling tail is  $K\approx 570$ (km/10\,kpc)$^2$. 
Interestingly, the best-fitting parameters for pure H and solar H/He ratio (with $Z \!=\! 0.01Z_{\odot}$) 
composition models are nearly identical. 
We have also checked how the data selection affects the best-fitting parameters. 
For the lower flux limit of $F_{\min}=0.5F_{\rm td}$, the parameters hardly change at all. 
However, taking $F_{\min}=0.1F_{\rm td}$ results 
in values for $\fedd$, $A$, and  $T_{\rm Edd,\infty}$  higher by 2--3 per cent for H and solar models. 
For He model, results are nearly independent of the data selection, because 
it describes the data somewhat better in a wider flux interval.

We further use the best-fitting  $A$ and $\fedd$ for the set of combined bursts (and $F_{\min}=F_{\rm td}/e$)  
to constrain the NS mass and radius. 
Pure helium models  can be rejected just from the value of  $T_{\rm Edd,\infty}$ (see Table~\ref{tab:results} and the 
solid red curve in Fig.~\ref{fig:mr_hard}(a)), because they either give a mass lying  below the mass-shedding limit or 
extremely high masses of $>4\msun$. The solar composition models 
predict $M$--$R$ dependence obtained from $T_{\rm Edd,\infty}$ that nearly coincides with  the mass-shedding limit 
(see solid blue curve in Fig.~\ref{fig:mr_hard}(a)), restricting  the hydrogen mass fraction to $X\gtrsim0.7$. 
Because of the nature of the companion star \citep{Wachter02}, there is no reason 
to take the hydrogen fraction below the solar value, therefore 
we assume a uniform distribution of $X$ in the interval from 0.74 to 1. 
For the values of $\fedd$ and $A$ for different $X$ we use linear interpolation between the 
corresponding values for $X=0.74$ and $X=1$. 
We further assume a  probability distribution function for the distance $D$ to \source\ 
to follow a Gaussian with the mean and standard deviation of 5.8 and 2.0~kpc, respectively 
(see Section~\ref{sect:distance}). 
Using Monte-Carlo simulations, we simulate $D$ and $X$
and convert a distribution of $A$ and $\fedd$ (obtained with a bootstrap) 
to the distribution of $M$ and $R$ (see \citealt{SPRW11} and Appendix~\ref{app:A}). 
We reject the solutions below the mass-shedding limit and 
with  $M<1.2\msun$ to be consistent with the NS formation scenarios 
\citep{WHW02} and the minimum observed pulsar masses  \citep{Kiziltan13}. 
We also reject solutions with $M>3\msun$, because there are no modern 
equation of state that support such massive NS. 
The  resulting banana-like contours (see Fig.~\ref{fig:mr_hard}(a)) 
are very much elongated along the curves of constant Eddington temperature $T_{\rm Edd,\infty}$,
which is just a result of a large uncertainty in distance. 
The width of the banana is defined by the errors in $T_{\rm Edd,\infty}$ and by the width of the distribution of $X$. 
The NS radius is constrained above 14 km (at 90 per cent confidence) 
independently of the metal abundance for NS masses in the range 1.2--2.4$\msun$. 
Note, however, that the radius becomes about 1 km 
smaller if we use $F_{\min}=0.1F_{\rm td}$ (see red dashed contours in Fig.~\ref{fig:mr_hard}(a)).
Thus the conservative lower limit for $R$ is 13~km. 
Our results are consistent with the stiff equation of state of cold dense matter that 
also has support from the recent observations of NS with $M\approx2\msun$ \citep{Demorest10,Antoniadis13}. 

We see that one branch of the solutions (high mass--small radius) corresponds to a larger gravity  than was 
assumed in the fits ($\log g=14.0$). Taking an atmosphere model for  $\log g=14.6$, does not 
affect at all the solutions for  $F_{\min}=F_{\rm td}/e$, but 
for $F_{\min}=0.1\,F_{\rm td}$ (red contours) it shifts the contours (upper left branch) down by 0.2$\msun$.  

We note that for the best-fitting $F_{\rm Edd,-7}=1.13$ and $A=0.137$--$0.138$ 
the solution for $M$ and $R$ exists only for $D_{10}$ below the upper limits 
$D_{10,\max}=0.45$ and 0.50 for H ($X=1$) and solar ratio H/He  ($X=0.74$) atmosphere, respectively (see Equation (\ref{eq:dmax})). 
At the maximum possible distance, the solution has to lie at the turning point (furthest from the origin of coordinates) 
of the curve of constant $T_{\rm Edd,\infty}$ on the $M$--$R$ plane. 
For a given distance $D<D_{\max}$, there are 
two solutions for $M$ and $R$ corresponding to $u_-$ (lower) and $u_+$ (upper) branches of solutions (\ref{eq:u_mr}).   
They can be found by substituting into Equations~(\ref{eq:R_Tedd}) 
the solutions obtained via Equations (\ref{eq:def_alpha}) and (\ref{eq:u_mr}) using the 
observables $A$ and $\fedd$ as well as the distance $D$. 
Restriction on  the NS mass $M>1.2\msun$ thus puts the lower limit on the distance $D_{10}>D_{10,\min}=$0.3 and 0.32 
for $X=1$ and 0.74, respectively. 
The presented error contours are barely consistent with the existing theoretical $M$--$R$ relations
and  at higher masses they are deviating even more (see Fig.~\ref{fig:mr_hard}(a)). 
Thus, it is likely that the NS mass in \source\  is not much larger than the typically measured 1.4--1.5$\msun$. 
For $M\in[1.2,1.5]\msun$ the distance has to lie in a rather narrow range between 3.1 and  3.7~kpc. 
 
If instead we follow the assumption of \citet{GO10} and introduce a sharp cut in the distance 
distribution at $D_{\min}=3.9$~kpc, the size of the contours in Fig.~\ref{fig:mr_hard} will be significantly reduced. 
For example, for  $X=1$ the contours will close at $M>1.9\msun$, 
which results in $R>15$~km (for $M<2.4\msun$), while for  $X=0.74$ 
similar constraints are $M>1.6\msun$ and $R>16$~km. 
It is clear that such a cut in $D$ would not produce realistic results for these bursts.

We note here that all the constraints obtained here are based on NS atmosphere model for non-rotating stars. 
Because \source\ rotates 620 times a second, the shape 
of the NS is distorted and the emission cannot possibly be spherically symmetric. 
Rapid rotation also boosts radiation emitted along the equatorial plane and hardens the spectrum. 
Including effects of rapid rotation would reduce the radius of the non-rotating NS determined from the 
cooling tail method by about 10 per cent  depending on the inclination (V. Suleimanov et al., in preparation).

\subsection{Comparison to the soft-state bursts and the touchdown method} 
\label{sec:soft}

Let us now take a look at the soft-state bursts. 
Because the evolution of $K^{-1/4}$ with flux does not follow the predicted theoretical 
dependence, this theory cannot be used  to get $f_{\rm c}$ and, therefore,
it is meaningless to use these data to determine NS parameters 
(using the cooling tail  or any other method). 
However, to demonstrate the main difference in the NS mass-radius constraints 
from the hard- and the soft-state  bursts, we apple 
the touchdown method as was done for \source\ by  \citet{GO10}.

\begin{figure*}
\centering 
\includegraphics[width=7.cm]{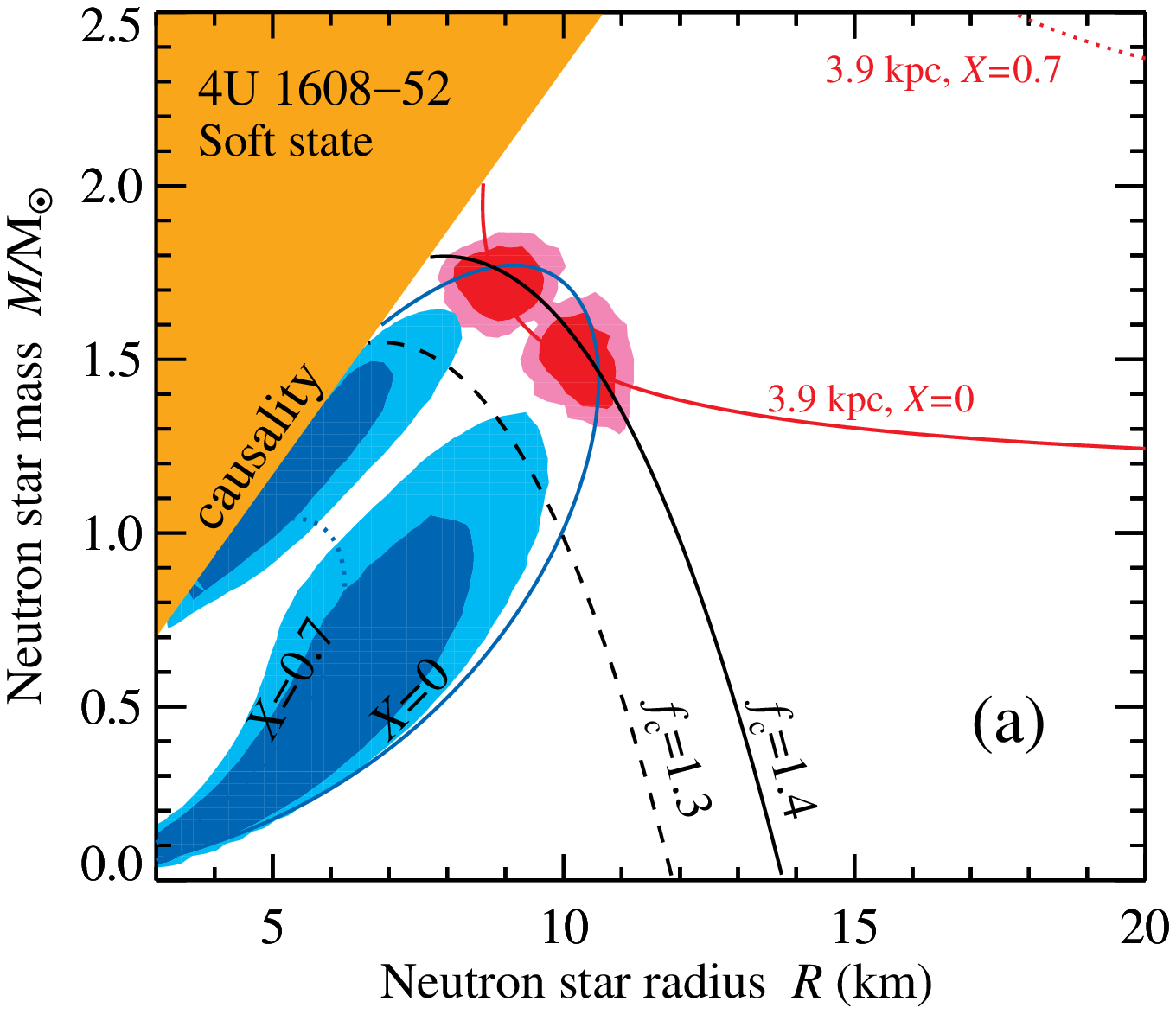}
\hspace{0.5cm}
\includegraphics[height=8.7cm,angle=90]{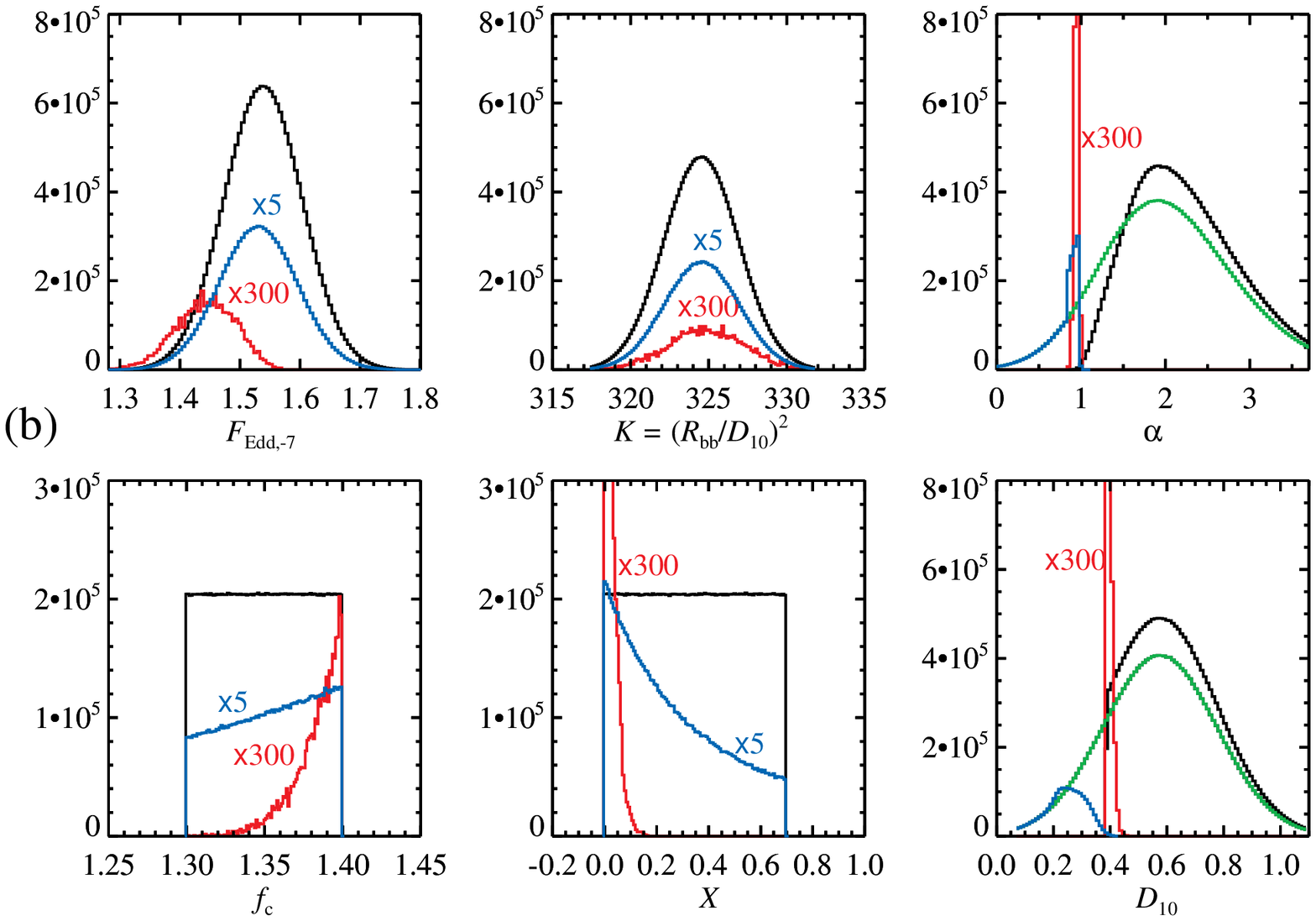}
\caption{(a) Mass--radius constraints from the analysis of the soft-state 
bursts from \source.
The lower left dark and light blue contours (for 68 and 90 per cent confidence limits) 
represent the constraints under an assumption of the normal distribution of the distance, 
while the  red and pink contours  (for 68 and 90 per cent confidence limits) 
correspond to a similar analysis with the cut in the distance distribution at 3.9~kpc \citep{GO10}. 
Both constraints are unreliable, because the soft-state bursts do not follow the theoretical models
these results are based on. 
The red curves show the constraints from the value of $\fedd$ and 
the blue curves give constraints from the values of $T_{\rm Edd,\infty}$ assuming  $f_{\rm c}=1.4$.  
The solid curves correspond to pure He case, $X=0$, and the dotted curves are for $X=0.7$. 
The black curves give the constraint from $K$ taking $f_{\rm c}=1.4$ (solid) and $f_{\rm c}=1.3$ (dashed) 
for  $D=3.9$ kpc. We see that only when simultaneously $X\approx0$, $f_{\rm c}\gtrsim 1.4$ and $D\lesssim4$ kpc,
the curve do cross and a solution exists. 
(b) Corresponding distributions of the parameters. 
The Monte-Carlo simulations were run with  $2\times10^7$ points.  
The black  histograms are the prior parameter distributions used by \citet{GO10}, 
with the distance distribution having a cutoff at 3.9 kpc. 
The green histograms give the distributions of $D$ and $\alpha$ without such a cutoff. 
The red histograms (multiplied by 300) are the posterior distributions of parameters that 
give a physical solution (see Appendix~\ref{app:A}) for the first case, while 
the blue histograms are similar distributions for the normal distance distribution.  
}	
\label{fig:m_r_soft}
\end{figure*}

First, for the touchdown method, we need to find the black-body normalization in the cooling tail. 
Looking at Fig.~\ref{fig:K14_f}(b), we see that the typical value of $K^{-1/4}$ is about 0.23, 
which translates to $K\approx350$.
Second, the flux at touchdown (which is \textit{assumed} to be equal to the Eddington flux $\fedd$)
for most bursts is between 1.0 and 2.0, with the average of about $1.6\times 10^{-7}$~erg~cm$^{-2}$~s$^{-1}$ 
(see stars in Fig.~\ref{fig:K14_f}(b) and Table~\ref{tab:bursts}).  
These values are  similar to those determined by \citet{GO10}:  
$\fedd = (1.541\pm0.065)\times 10^{-7}$~erg~cm$^{-2}$~s$^{-1}$ and $K = 324.6\pm2.4$ in the cooling tail,
which we adopt for easier comparison.\footnote{  
They have used bursts 1 and 8 from Table~\ref{tab:bursts} for determination of the touchdown flux and 
bursts 6, 7, 8 and one non-PRE burst to measure the blackbody normalisation of the cooling tail.}
They correspond to $T_{\rm Edd,\infty} = 2.14\times10^7$~K and the maximum 
possible distance $D_{10,\max} =0.405$ at $f_{\rm c}$=1.4, $X$=0  and central values for 
$\fedd$ and $K$ (see Equation~(\ref{eq:dmax})). 
Taking $f_{\rm c}$ smaller and $X$ larger decreases  $D_{\max}$ further.
We note here that in the hard-state bursts 
$K$ is larger by a factor of 1.75, $\fedd$ is smaller by 50 per cent and $T_{\rm Edd,\infty}$ is smaller by $\sim$30 per cent.  
 
We now can use these observables to  obtain the NS mass and radius distribution.  
We follow here the assumptions by \citet{GO10}: we take a uniform distribution of $f_{\rm c}$ 
between 1.3 and 1.4 (although the actual value for He atmosphere is closer to 1.5, see \citealt{SPW12}), 
and assume a uniform distribution of the hydrogen fraction $X$ between 0 and 0.7 
(this is also questionable, because the companion star is likely hydrogen rich, see \citealt{Wachter02}). 
However, for the distance $D$  distribution, we take a Gaussian with the mean 5.8~kpc and $1\sigma$ of 2~kpc. 
These distributions are shown by black (or green for $D$ and parameter $\alpha$,  see Appendix~\ref{app:A}) histograms in Fig.~\ref{fig:m_r_soft}(b). 
The posterior distributions of parameters that give a physical solution for $M$ and $R$ 
are shown by blue histograms. 
The solution exists for about 10 per cent of all parameters (just because of the constraint $D<D_{\max}\approx 4$~kpc). 
The resulting NS mass and radius are very small (see blue contours in Fig.~\ref{fig:m_r_soft}(a)) 
with the best-fitting $R$ being below 8 km, because the peak in the distance distribution is at  $\sim$2~kpc. 
If we take a distribution of $X$ extending to 1, the solutions will be even more extended towards lower radii. 
On the other hand, extending $f_{\rm c}$  towards 1.5 leads to extension of the contours to  radii up to 12 km and masses 
to $2\msun$. 
Cutting the NS mass distribution at $M>1.2\msun$ leaves two separated regions: 
$R\in[5,8]$~km for larger mass $M\in [1.2,1.6]\msun$ and $R\in[7,9]$~km for $M\in [1.2,1.35]\msun$.

These results are very different from $M=1.74\pm0.14\msun$ and $R=9.3\pm1.0$~km 
obtained by  \citet{GO10} from the same data.\footnote{The errors on the NS mass 
and radius obtained by  \citet{GO10} are extremely small, i.e. much smaller than the uncertainties in the distance; 
they just reflect the statistical errors on $K$ and $\fedd$.} 
The only difference in our approach is that we did not cut the distance distribution. 
To illustrate this, we now apply such a cut, leaving only $D>3.9$~kpc. 
The posterior distributions of parameters that give a physical solution 
are shown by red histograms in Fig.~\ref{fig:m_r_soft}(b). 
Only a fraction of about $6\times 10^{-4}$ of parameters $A$, $\fedd$ and $D$  
from the assumed prior distributions produce a physical solution for the NS mass and radius  
(note that in Fig.~\ref{fig:m_r_soft}(b) the posterior distributions are multiplied by a factor of 300). 
We see that  the posterior distribution of $\fedd$ is strongly skewed towards smaller flux by $\sim1.5\sigma$, 
$f_{\rm c}$ distribution is skewed towards larger values and the allowed chemical composition is nearly pure helium with $X<0.1$ 
(which contradicts the nature of the companion, \citealt{Wachter02}), 
the distribution of parameter $\alpha$ (see Equation (\ref{eq:def_alpha})) and 
the distance distribution  are nearly a $\delta$-function (because all solutions have to lie between the cutoff at 3.9~kpc 
and $D_{\max}=4.05$~kpc).  
Thus it is clear that the values for the best-fitting $M$ and $R$ and their small errors 
(see red contours in Fig.~\ref{fig:m_r_soft}(a)) are fully determined by 
an unrealistic assumption of the sharp cutoff in the distance distribution  and a cut in the distribution of $f_{\rm c}$ at 1.4. 

The problem that a very small fraction of the parameter space gives a physical solution 
was noticed previously by  \citet{SLB10}. 
As a solution, they suggested to relax an assumption that the touchdown moment corresponds to the photosphere being 
at the NS surface. 
They, however, adopted all the observables given in \citet{GO10} and assumed the same distance distribution with 
the unphysical cutoff.  
We showed here that none of the parameters presented  by \citet{GO10} (i.e. $K$, $\fedd$, $X$ or $f_{\rm c}$) 
is actually correct and the  combination of these parameters with the cut in the distance distribution was 
responsible not only in small errors and the values of the determined $M$ and $R$ but also 
in the smallness of the parameter space giving solutions. 
Thus, the solution to the problem actually lies in picking up different bursts that do follow the theoretical atmosphere models.

Similar problems  appear in the analysis of a number of other sources, e.g., 
4U\,1820--30  \citep{OGP09} and EXO\,1745--248 \citep{GWCO10}. 
All these bursts, that were used to determine the NS mass and radius, happened 
at high accretion rate, and did not follow the spectral evolution predicted by the theoretical atmosphere models. 
The errors on mass and radius in these sources were also very small,   
again due to  an  artificial sharp cut in the distance distribution, which 
fully determined the outcome \citep[see ][for a discussion]{SPRW11}. 
In recent papers, on KS\,1731$-$260  \citep{OGG12} and SAX\,J1748.9$-$2021 \citep{GO13}, 
the  cuts in the distance distribution were not applied, and as a result, the $M$-$R$ contours extended to lower left corner, similarly 
to our blue contour in Fig.~\ref{fig:m_r_soft}(b), which then could be reduced by the condition $M>1.2\msun$. 
For both sources, however, the analysed bursts again did not follow theoretical evolution and 
thus the results are dubious.

Because the cooling tracks for the soft-state bursts do not  follow theoretically predicted behaviour for a passively cooling NS, 
it is not surprising that the results for the NS mass and radius determination 
are very much different for the soft- and the hard-state bursts. 
There are also physical reasons that explain the difference. 
First, at  high accretion rate, the accretion disc blocks half of the NS. 
Second, the colour-correction factor in the soft-state bursts is expected to be much larger than the usually assumed $\sim$1.4, 
if a significant part of the burst radiation has to pass through the rapidly rotating spreading layer above the  NS surface \citep{IS99},
which has a reduced effective gravity due to the centrifugal support, leading to a flux  through the atmosphere that  
is close to the local Eddington limit and has a high colour correction $f_{\rm c} \approx 1.6-1.8$ \citep{SulP06}.
These effects together  naturally explain why the black-body normalisation is constant and much smaller 
during the cooling tail of the soft bursts \citep{SPRW11}. 
It is clear that the constancy of the  black-body area alone cannot be used as an 
argument in favour  of visibility of the whole star. 
Moreover, the constancy of the apparent area contradicts theoretical atmosphere models. 
The presence of the accretion disc also reflects a part of the burst radiation 
breaking the spherical symmetry, increasing the observed flux at some angles \citep{LS85} 
and leading to  a higher touchdown flux in the soft-state bursts.  

We conclude that as the variations in $K^{-1/4}$ as a function of flux 
do not follow theoretical predictions, these bursts cannot presently be used 
reliably to infer NS $M$ and $R$. 
 
\section{Summary}
\label{sec:conclusions}

We studied 21 PRE bursts from \source\ and found a clear dependency of the spectral evolution in 
the cooling tail of the bursts on the spectral state of the persistent emission prior to the burst. 
The same dependency can be also seen from various other model-independent parameters.
We showed that the bursts observed during the hard state at low accretion rates 
are consistent with theoretical predictions of the NS atmosphere models, while the spectral evolution of the 
soft-state bursts is inconsistent with the theory. 
Such a behaviour is ubiquitous in the atoll sources \citep[see][]{Kajava14}. 
This implies that the basic assumption of the passively cooling NS atmosphere breaks down in these bursts. 
We argue that only the hard-state bursts at persistent luminosities below a few per cent of Eddington 
with the colours of the persistent emission characterised by $1<S_z<2$ 
can be used in efforts of determination of the NS masses and radii from the thermal emission of X-ray bursts. 

We applied the most recent set of NS atmosphere models 
that account for Klein-Nishina reduction of electron opacity 
to the data of the hard-state PRE bursts of \source\ and obtained the Eddington flux and the apparent angular size of the NS.
From these values, we constrain the mass and radius of the NS.
Because of a large uncertainty in the distance, the solution lies along a long strip of constant Eddington temperature. 
For typically assumed and measured NS masses between 1.2 and 2.4$\msun$, the NS radius is strongly 
constrained to be above 13 km. 
These constraints are similar to that found for 4U~1724--307 \citep{SPRW11} and
supports a stiff equation of state for the cold, dense matter of the NS interior.  
These are also consistent with the recent discoveries of the two-solar-mass pulsars \citep{Demorest10,Antoniadis13}. 
We note, however, that the NS parameters are obtained assuming spherically-symmetric non-rotating NS, 
covered by an atmosphere with homogeneous distributions of the effective temperature, 
 surface gravity and  chemical composition, and described by a single plane-parallel model atmosphere.
Rapid rotation, however, breaks the symmetry and affects the NS shape, the observed flux and the colour temperature. 
A preliminary study (V. Suleimanov et al., in prep.)  indicates that for a NS in \source, which rotates at 620~Hz, 
the lower limit on the radius of the non-rotating NS may be reduced by as much as 10 per cent. 
In this case,  constraints on the NS radius  from X-ray bursts 
become consistent with those coming from nuclear physics  \citep{LS14}. 

Finally, we note that constraints on the NS masses and radii 
obtained from the soft-state and/or high-accretion rate bursts that do not follow theoretically predicted spectral evolution
\citep{OGP09,OGG12,GO10,GWCO10,GO13,SLB10,LS14}  have to be revisited.

\section*{Acknowledgments}
The research was supported by Academy of Finland grant 268740 (J.P.), 
the V\"ais\"al\"a Foundation (J.N., J.J.E.K.),  
the Finnish Doctoral Programme in Astronomy and Space Physics
and the Emil Aaltonen Foundation (J.J.E.K.). 
V.S. acknowledges funding from the German Research Foundation  (DFG) grant SFB/Transregio 7 ``Gravitational Wave Astronomy'' 
and the Russian Foundation for Basic Research (grant 12-02-97006-r-povolzhe-a). 
DKG acknowledges the support of an Australian Research Council Future Fellowship (project FT0991598). 
We also acknowledge the support of the International Space Science Institute (Bern, Switzerland), 
where part of this investigation was carried out.   
This research has made use of data provided by the High Energy Astrophysics Science Archive Research Center (HEASARC), 
which is a service of the Astrophysics Science Division at NASA/GSFC 
and the High Energy Astrophysics Division of the Smithsonian Astrophysical Observatory.




\label{lastpage}

\appendix

\section{The cooling tail method} 
\label{app:A}

For completeness we present  the cooling tail method \citep{SPRW11,SPW11} that 
is used in the paper to constrain the NS mass and radius. 
In order to achieve this goal, we need measurements of the apparent area of the NS and 
the Eddington flux. Knowing the distance to the source is also a prerequisite for accurate calculations.

We define the Eddington flux as 
 \be \label{eq:fedd}
\fedd= \frac {G Mc}{D^2 \kappa_{\rm e}(1+z)} ,  
\ee
where $\kappa_{\rm e}=0.2(1+X)$ cm$^2$g$^{-1}$ is the electron scattering opacity, 
 $X$ is the hydrogen mass fraction,  
$D$ is the distance to the source,  
\be
1+z=\left(1- u\right)^{-1/2} =  \left(1- 2GM/Rc^2\right)^{-1/2} 
\ee
is the surface redshift, $u=\rs/R$, $R$ is the NS stellar circumferential radius, 
$\rs=2GM/c^2$ is the Schwarzschild radius, and $M$ is the NS gravitational mass. 
Such definition of the Eddington flux explicitly assumes that the opacity is dominated  
by electron scattering and the cross-section is equal to the Thomson one. 
For the case of hot atmospheres of NS with the surface temperature of about 3--3.5 keV, 
which correspond to the Eddington flux at the NS surface for $\log g=14-14.5$, 
the Rosseland mean Compton scattering opacity is about 7--10 per cent lower that the Thomson one, 
resulting in an Eddington limit which is larger by the same amount \citep{SPW12}.
This immediately implies that  even if the Eddington limit is reached at the touchdown, 
the flux at this moment  has to be larger that $\fedd$, which is defined for Thomson opacity. 
 
At luminosities close to the Eddington, the spectrum of the NS is close to 
a diluted black body $F_E\approx w B_E(T_{\rm c}=f_{\rm c} T_{\rm eff}$, 
where $T_{\rm eff}$ and $T_{\rm c}$ is the effective and colour temperatures measured 
at the NS surface,   
 $f_{\rm c} $ is the colour-correction factor and $w$ is the dilution factor which  is close to $1/f_{\rm c}^{4}$. 
The observed bolometric black-body flux is 
\be
\label{eq:Fobs}
     F = \sigma_{\rm SB} T_{\rm bb} ^4  \frac{R_{\rm bb} ^2}{D^2}=\sigma_{\rm SB} 
     T_{\infty}^4 \frac{R_{\infty}^2}{D^2},  
\ee
where $R_{\infty}=R(1+z)$ is the apparent NS radius,  
$T_{\infty}$ is the redshifted effective temperature  and 
$T_{\rm bb} $ is the measured colour temperature. 
The temperatures are related as 
\be
\label{eq:tempdep}
    T_{\rm bb} = f_{\rm c}\ T_{\infty}= f_{\rm c}\ \frac{T_{\rm eff}}{1+z}= \frac{T_{\rm c}}{1+z}.
\ee
We then get the relation between the black-body normalisation and the NS radius
\be
\label{eq:rbbD2}
\frac{R_{\rm bb}^2}{D^2} =  \frac{R^2}{D^2} \frac{(1+z)^2}{f_{\rm c}^4} =  \frac{R_{\infty}^2}{D^2} \frac{1}{f_{\rm c}^4} ,
\ee
which can be easily transformed to \citep{Penninx89,vP90} 
\be \label{eq:KAfc}
K^{-1/4}=f_{\rm c} A,  \qquad A=(R_{\infty} [{\rm km}]/D_{10})^{-1/2} ,
\ee
where $K=(R_{\rm bb}\mbox{[km]}/D_{10})^2$ and $D_{10}=D/10\ \mbox{kpc}$. 
We can rewrite this as 
\be\label{eq:def_rinftyb}
R_{\infty} =  R  (1+z) = D_{10}  A^{-2}= D_{10} \sqrt{K} f_{\rm c}^2 \ \mbox{km}.  
\ee   
A combination of  $A$ and $\fedd$  gives a distance-independent quantity, 
the Eddington temperature:  
\be \label{eq:tedd}
T_{\rm Edd,\infty}  = \left(\frac{gc}{\sigma_{\rm SB} \kappa_{\rm e} }\right)^{1/4} \!\! \frac{1}{1+z}
= 1.14\times 10^8\, A\,F_{-7}^{1/4}\ \mbox{K} , 
\ee 
where $F_{-7} = \fedd / 10^{-7}$ erg cm$^{-2}$ s$^{-1}$.
This gives a parametric relation between radius and mass of the NS via compactness $u$: 
\bea \label{eq:R_Tedd}
R & =&   \frac{c^3 u (1-u)^{3/2}}{2  \kappa_{\rm e}  \sigma_{\rm SB} T_{\rm Edd,\infty}^4 }= 1188\frac{u (1-u)^{3/2}}{(1+X)\, T_{\rm Edd,\infty,7}^4} \ \mbox{km}, \nonumber \\
m & \equiv & \frac{M}{\msun} = u \frac{R}{2.95\ \mbox{km} } ,
\eea
where $T_{\rm Edd,\infty,7}=T_{\rm Edd,\infty}/ 10^{7}\mbox{K}$. 
It is easy to see that for the same $T_{\rm Edd,\infty}$ and same compactness $u$, the 
NS radius is two times larger for helium model ($X=0$) than for hydrogen model ($X=1$).
 
Expression (\ref{eq:KAfc}) gives the basis for the cooling tail method \citep{SPRW11,SPW11}. 
From the observations we obtain the relation between $K^{-1/4}$ and flux $F$, which 
is fitted with the theoretical dependence $A\,f_{\rm c}(F/\fedd)$ with 
the free parameters being  $\fedd$ and normalisation $A$. 
This is different from the touchdown method  that was used by   
\citet{OGP09} and \citet{GO10,GWCO10}, who  
assume that the touchdown flux is equal to the Eddington flux, 
take some value for black-body normalisation $K$  in the cooling tail and assume a value for $f_{\rm c}$. 
The cooling tail method uses the information from the whole cooling tail (not only two numbers), 
actually allows to check (e.g. in terms of $\chi^2$) whether the spectral evolution 
in the cooling tail is consistent with the NS atmosphere models, does 
not require any assumption for the colour-correction factor in the cooling tail and 
is not based on the assumption of equality of the touchdown and the Eddington flux. 
Of course, the theoretical dependence of $f_{\rm c}$ still depends 
on the chemical composition of the atmosphere and the gravity. 
The gravity can be varied to get a self-consistent solution for the mass and radius. 

Once $A$ and $\fedd$  are known and some value for the distance is assumed, 
it is easy to compute the resulting NS mass and radius. 
Radius at infinity is found via Equation (\ref{eq:def_rinftyb}).
From the Eddington flux we can find the quantity 
\be\label{eq:def_c}
C \!=\!   \frac{\rs }{1+z} = 0.4 \frac{\fedd D^2(1+X)}{c^3} = 14.1 (1+X) D_{10}^2 F_{-7}\ \mbox{km}. 
\ee  
Combining Equations (\ref{eq:def_rinftyb}) and (\ref{eq:def_c}), we 
can define a dimensionless quantity 
\be 
\label{eq:def_alpha}
\alpha \equiv 4C/R_{\infty} = 4 u(1-u) = 56.5 (1+X)  F_{-7} {A}^{2} D_{10}   . 
\ee 
Equation (\ref{eq:def_alpha}) can be solved for $u$: 
\be \label{eq:u_mr}
u = u_{\pm}= \frac{1}{2 } \left( 1\pm \sqrt{1- \alpha } \right) . 
\ee
The solution exists only if $\alpha\leq 1$, 
which can be translated to the upper limit on the distance  
\be \label{eq:dmax}
D_{10} \! \leq \!  D_{10,\max} \!=\! \frac{1.77\times 10^{-2}}{(1+X)A^2 F_{-7}} 
\!=\!  1.77\times 10^{-2}\frac{K^{1/2} f_c^2}{(1+X)F_{-7}}. 
\ee

Solutions (\ref{eq:u_mr}) can be rewritten for the redshift factor  
\bea
(1+z)_- &= & \left( \frac{2}{1 + \sqrt{1- \alpha} } \right)^{1/2}, \\
(1+z)_+&= & \frac{2}{\alpha^{1/2} }   \left( \frac{1 + \sqrt{1- \alpha}}{2} \right)^{1/2} . 
\eea 
And finally the NS mass and radius can be obtained 
\bea \label{eq:u_mr_final}
m&=& \frac{C}{2.95\ \mbox{km}} (1+z) = \alpha (1+z) \frac{R_{\infty}}{11.8\ \mbox{km}}, \nonumber \\
R &= & R_{\infty}/(1+z). 
\eea
From given prior distributions of  $A$, $\fedd$  and $D$, using Equations (\ref{eq:def_c})--(\ref{eq:u_mr_final}) we 
can simulate the posterior distribution of $M$ and $R$.

\end{document}